\documentclass[]{JHEP3}
\usepackage{cite}
\usepackage{pinlabel}
\usepackage{subfigure}
\usepackage{graphicx}
\usepackage{amsmath}
\numberwithin{equation}{section}

\preprint{Cavendish-HEP-2011/24}
\title{Effect of spin-3/2 top quark excitation on $t\bar{t}$ production at the LHC}
\author{W.J.~Stirling, E.~Vryonidou\\Cavendish Laboratory, J.J. Thomson Avenue, Cambridge CB3 0HE, UK}

\abstract
{A spin-3/2 top quark excitation originating in string realisations of the RS scenario has an effect on the production of top-antitop pairs in hadron-hadron collisions. We study the additional contribution of this state to the cross section for the LHC. We comment on the prospect of discovery or exclusion for a range of masses and couplings at both 7 and 14~TeV and show that the reach extends significantly with increasing centre-of-mass energy. Results are compared with the effect of a hypothetical spin-1/2 excited top as well as other BSM scenarios which predict the presence of higher dimension effective operators. We also calculate distributions for several observables which can be used to distinguish between the different scenarios in the presence of a deviation from the SM prediction for the $t\bar{t}$ cross section. }

\begin{document}

\tableofcontents

\section{Introduction}
Top quark pair production is expected to be an important window to New Physics (NP) at the Large Hadron Collider (LHC). The top quark, due to its large mass, is expected to play a special role in electroweak symmetry breaking and its coupling to new physics is expected to be large. The LHC has already produced the first quantitative results in top quark physics and more are expected with more data being accumulated. The main top quark production mechanism at the LHC is pair production from gluon-gluon fusion. The first measurements of the top pair cross production section at the LHC using different detection channels have been published by ATLAS \cite{Aad:2011yb} and CMS~\cite{Chatrchyan:2011nb,Chatrchyan:2011ew,Chatrchyan:2011yy}. 

Several Beyond the Standard Model (BSM) scenarios predict deviations from the Standard Model (SM) prediction for top pair production. These deviations manifest themselves as changes in both the shapes of distributions of appropriate kinematical variables as well as in the value of the total cross section. Results from the Tevatron have set constraints on NP effects in $t\bar{t}$ production and in particular on the existence of resonances decaying to top pairs~\cite{Abazov:2008ny,Aaltonen:2009qu,Aaltonen:2011ts}. However as the top pair production at the Tevatron comes mostly from quark-antiquark annihilation, NP contributions to $gg\rightarrow t\bar{t}$ remain largely unconstrained. The LHC is expected to shed more light on these unexplored regions. Experimental searches for NP in pair production are already underway, with both ATLAS and CMS having preliminary results on NP resonances decaying into top-antitop pairs \cite{ATLASres,CMSres}.

On the phenomenological side, several studies of NP in the top quark sector have been performed for both resonant and non-resonant contributions to the top pair production. Resonant models discussed in the literature include Z', KK gluons, axigluons, RS gravitons. Recently there has been a lot of interest in BSM scenarios affecting $t\bar{t}$ production in the light of the forward-backward asymmetry anomaly observed by CDF~\cite{Aaltonen:2011kc} and later confirmed by D0~\cite{Abazov:2011rq}. New physics models have been recently proposed or revisited to try to explain the discrepancy between the SM prediction and the experimental results. For resonant scenarios much work has been devoted to studying the different spin resonances and to identifying ways to distinguish between them, see for example \cite{Frederix:2007gi,Barger:2006hm}. Non-resonant physics studies have focused on the use of effective theory Lagrangians involving the presence of higher-dimension operators. The origin of these operators is related to theories of top quark compositeness~\cite{Kumar:2009vs} and resonance models where the exchanged particle is very heavy and therefore integrated out, leading to higher-dimension effective operators suppressed by the mass of the resonance, e.g. \cite{Jung:2009pi}. Of all possible operators a set  consistent with the SM symmetries is extracted and studied in the literature. These operators lead to anomalous top interactions, modifying the top-antitop-gluon vertex and introducing new interactions with the light quarks. For a comprehensive study of dimension-six operators: both four-fermion operators and the chromomagnetic operator and their effect on $t\bar{t}$ pair production, see \cite{Degrande:2010kt} and references therein.

Another interesting NP scenario having an impact on top pair production is the existence of a spin-3/2 top excitation. Excited quarks and leptons have been studied in the past within the context of compositeness of quarks and leptons, which predicts both spin-1/2 and spin-3/2 excited states. In particular spin-3/2 excitations have been studied in \cite{Burges:1983zg,Kuhn:1984rj,Kuhn:1985mi,Moussallam:1989nm,Almeida:1995yp,Dicus:1998yc,Walsh:1999pb,Cakir:2007wn}. The interactions of the new states are described by effective Lagrangians where higher-dimension terms are suppressed by the compositeness scale. Another motivation for a spin-3/2 top excitation arises in the context of higher-spin Regge excitations. The scenario we are considering in this paper is the existence of a spin-3/2 top excitation as proposed in \cite{Hassanain:2009at} within a string theory inspired model of warped extra dimensions. This new state is thought to be the lightest in a tower of Regge excitations. In \cite{Hassanain:2009at} the single and pair production cross sections of this new state are studied at the Tevatron and the LHC. Here we will focus on virtual effects of this excited state and in particular its impact on $t\bar{t}$ production. This additional contribution arises from the mixing of this new state with the SM top through the emission of a gluon described by a dimension-five term in the effective Lagrangian.

In this study we first reintroduce the spin-3/2 excited top and its interactions, reviewing some results already in the literature in Section 2. In Section 3 we study the impact of this additional state on top pair production, investigating the importance of mass and couplings at the LHC. A comparison with the effect of a hypothetical spin-1/2 top excitation  is presented in Section 4. In Section 5 we also compare with the contribution of the dimension-six chromomagnetic operator as an example of other non-resonant NP physics  effects in $t\bar{t}$ production, before we conclude in Section 6.

\section{Spin-3/2 top excitation}  
In string realisations of the Randall-Sundrum scenario, higher-spin Regge excitations of the SM particles arise. In general these are expected to be heavier than the Kaluza-Klein excitations of SM particles which arise when fermions are allowed to propagate in the bulk~\cite{Gherghetta:2000qt}. However it is argued in \cite{Hassanain:2009at} that individual light string excitations can be lighter than the KK modes and that the lightest of these higher-spin excitations is the spin-3/2 excitation of the right-handed top quark. The interactions of this spin-3/2 state are described by an effective theory Lagrangian respecting the gauge symmetries of the SM.  
The spin-3/2 top is represented by a Rarita-Schwinger field~\cite{Rarita:1941mf}. The free Lagrangian is:
\begin{equation}
\mathcal{L}=i\bar{\psi}_{\mu}\gamma^{\mu\nu\rho}D_{\nu}\psi_{\rho}+M\bar{\psi}_{\mu}\gamma^{\mu\rho}\psi_{\rho}
\label{free}
\end{equation}
and leads to the propagator:
\begin{equation}
P^{\mu\nu}=\frac{1}{p^2-M^2}\big[-(\displaystyle{\not}{p}+M)(\eta^{\mu\nu}-\frac{p^{\mu}p^{\nu}}{M^2})-\frac{1}{3}(\gamma^{\mu}+\frac{p^{\mu}}{M})(\displaystyle{\not}{p}-M)(\gamma^{\nu}+\frac{p^{\nu}}{M})\big].
\label{prop}
\end{equation}

\begin{figure}[h]
\begin{minipage}[b]{0.5\linewidth}
\centering
\includegraphics[trim=1cm 0 0 0,scale=0.6]{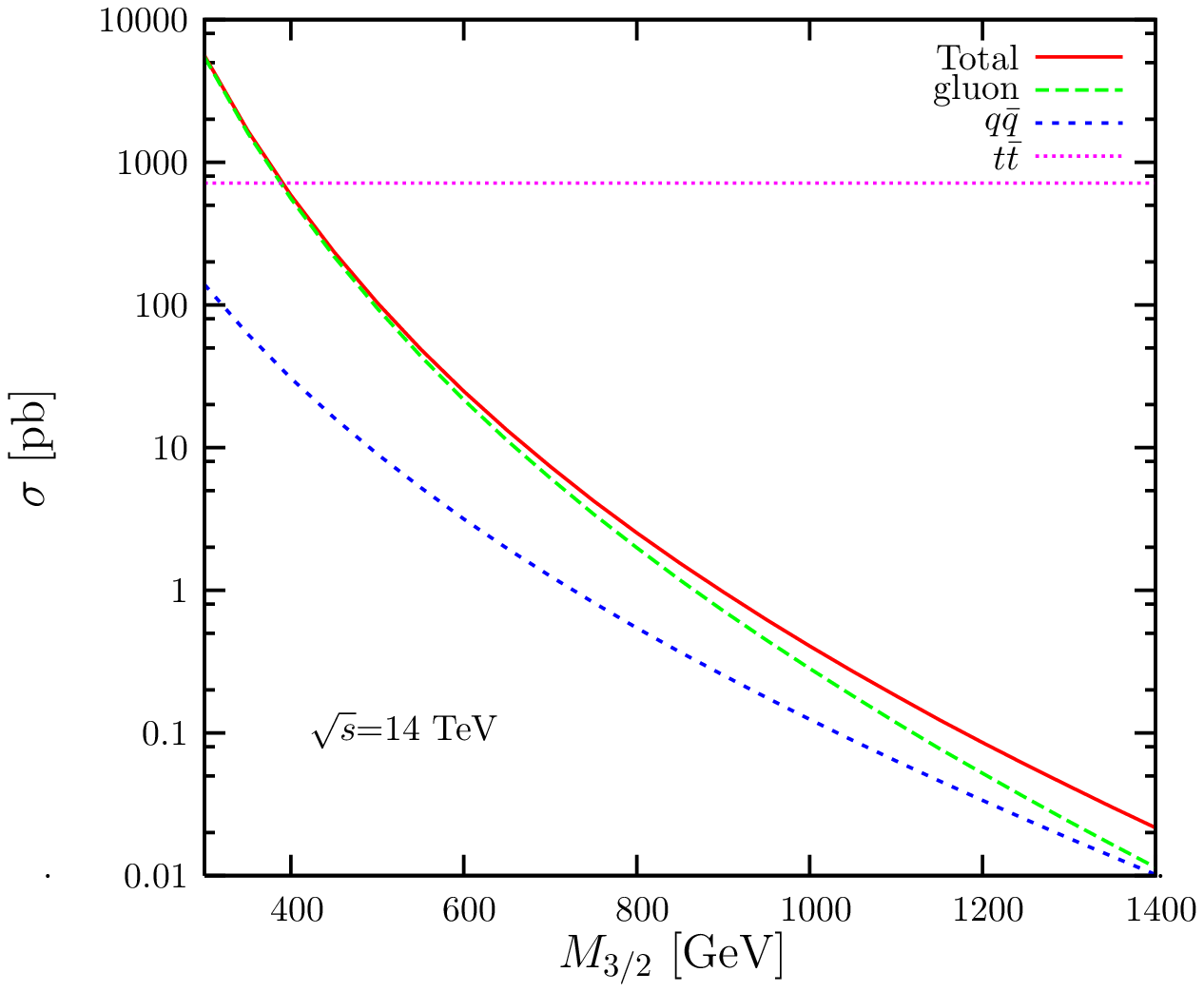}
\caption{Pair production cross section of spin-3/2 tops at the LHC (14~TeV) as a function of their mass.}
\label{pair}
\end{minipage}
\hspace{0.5cm}
\begin{minipage}[b]{0.5\linewidth}
\centering
\includegraphics[trim=1.3cm 0 0 0,scale=0.6]{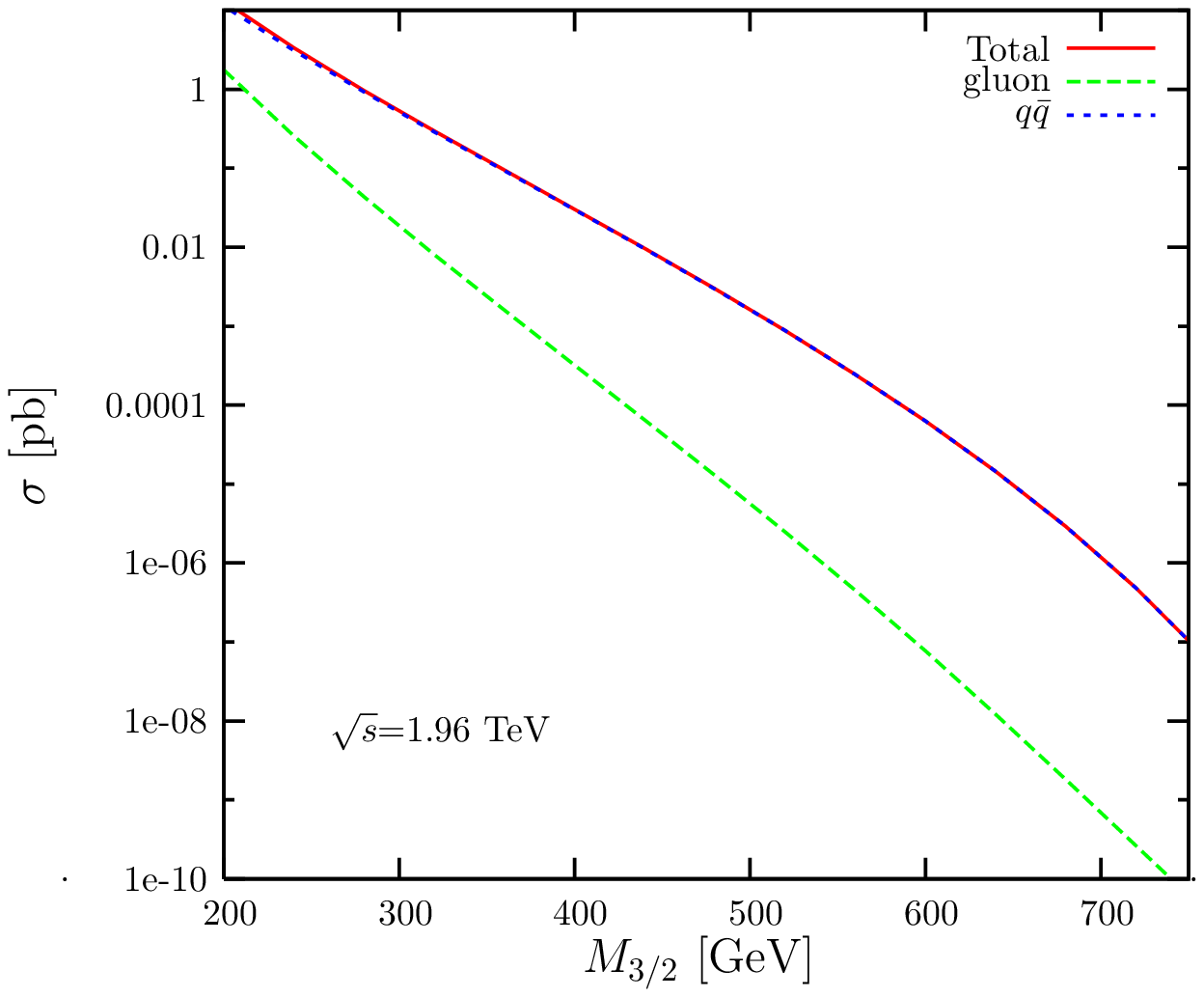}
\caption{Pair production cross section of spin-3/2 tops at the Tevatron (1.96~TeV) as a function of their mass.}
\label{pairtev}
\end{minipage}
\end{figure}

The first term in Eq.~\ref{free} leads to an interaction of the excited top with the gauge fields. Identical Lagrangians have been used in the literature to study spin-3/2 excited quarks and leptons. We derive the relevant Feynman rules and as a first check we attempt to reproduce the results for pair production of the excited quarks which exist in the literature. Pair production of spin-3/2 excited quarks was studied in \cite{Moussallam:1989nm}, \cite{Dicus:1998yc} and \cite{Hassanain:2009at}. The first two consider pair production of a generic excited spin-3/2 quark while the third studies this in the context of spin-3/2 Regge excitations of the right-handed top quark. The relevant Lagrangian terms are identical in both cases. 
However the numerical results of papers \cite{Hassanain:2009at} and \cite{Dicus:1998yc} for the total cross section at the LHC seem to be in disagreement by about an order of magnitude. We reproduced the analytical expressions for the partonic cross sections given in \cite{Dicus:1998yc}. We show our results for the pair production cross section as a function of the mass of the spin-3/2 top in Fig.~\ref{pair} using MSTW2008LO \cite{Martin:2009iq} PDFs and setting the factorisation and renormalisation scales equal to the mass of the excited top. For comparison with Fig.~6 of \cite{Hassanain:2009at} we also show the corresponding cross section at the Tevatron in Fig.~\ref{pairtev}. In the plots we decompose the cross sections into contributions from quark-antiquark and gluon-gluon scattering. At the Tevatron the cross section is completely dominated by quark-antiquark annihilation while at the LHC the dominant contribution comes from gluon-gluon scattering, in a similar way as in SM $t\bar{t}$ production. We note that the agreement with \cite{Hassanain:2009at} is much better for the Tevatron plot. The origin of the difference may be partly caused by the choice of different PDF sets, the value of $\alpha_s$  and possibly the choice of scale. Calculating the cross section using CTEQ6L PDFs~\cite{Pumplin:2002vw} significantly improves agreement. For the calculation with CTEQ6L we used a leading-order (LO) formula for $\alpha_s$. The remaining difference is eliminated if we choose a different scale $Q=\sqrt{\hat{s}}$. We note here that these LO results are strongly dependent on the choice of scale, with a cross-section variation of more than a factor of three for the standard scale variation $m_t/2<Q<2m_t$. 

\begin{figure}[h]
\centering
\includegraphics[scale=0.6]{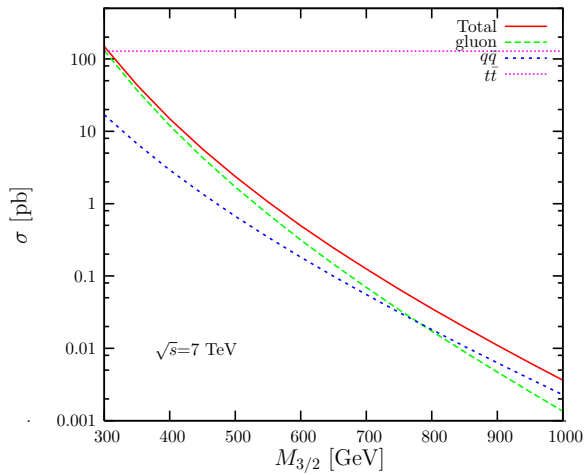}
\caption{Pair production cross section of spin-3/2 tops at the LHC (7~TeV) as a function of their mass.}
\label{pair7}
\end{figure}

In the graph we also show the value of the LO $t\bar{t}$ cross section at the LHC again using the MSTW2008LO PDF set. The corresponding value for the Tevatron is 7~pb which would lie at the very top of Fig.~\ref{pairtev}. The corresponding results for the current LHC running Centre-of-Mass (CoM) energy of 7~TeV are shown in Fig.~\ref{pair7}.

\begin{figure}[h]
\centering
\includegraphics[trim=1.5cm 0 0 0 ,scale=0.7]{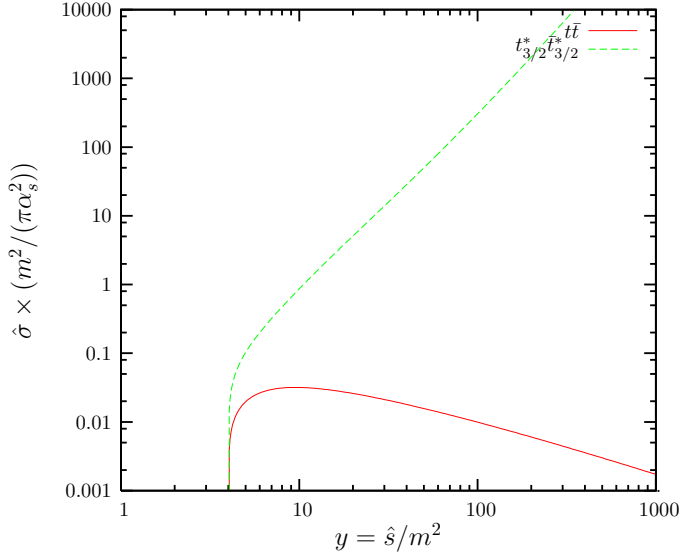}
\caption{Partonic cross section for the gluon contribution to pair production of $t$ and $t^*_{3/2}$.}
\label{shat}
\end{figure}

We comment here that as shown in Figs.~\ref{pair} and \ref{pair7} the cross section for the production of a spin-3/2 excited top pair is much higher than the production of a SM top pair of the same mass. This is a consequence of the higher spin which gives extra momentum factors in the propagator of Eq.~\ref{prop} and similarly in the expression for the spin sum. The same conclusion can also be drawn by considering the partonic cross section $\hat{\sigma}$ for gluon-gluon scattering as a function of the variable $y=\hat{s}/m^2$. This is shown in Fig.~\ref{shat}. We see as expected that the cross section for $t^*_{3/2}$ is rising rapidly in a clear manifestation of unitarity violation, while the partonic cross section for spin-1/2 pair production falls to zero at very high energies. In \cite{Hassanain:2009at} it is suggested to introduce a cut-off $\Lambda=7M_{3/2}$ to prevent the calculation from violating unitarity. This will be further discussed in the following Sections. Introducing this cut-off for pair production effectively means constraining the calculation in values of $y$ in Fig.~\ref{shat} below 49. The impact of this cut-off on Figs.~\ref{pair}-\ref{pair7} is very small for all but very low masses as the hadronic cross section is a rapidly falling function of the partonic CoM energy due to the PDF suppression of high momentum fractions. The decrease in the cross section for a given mass is more important at the LHC (14~TeV) where the available energy is larger. The threshold functional behaviour near $y=4$ is the same for both cases with the cross section rising as $\sqrt{y-4}$. This is a characteristic of the fermionic nature of the produced particles.

\section{Effect on top quark pair production}
\begin{figure}[h]
\centering
\includegraphics[scale=0.7]{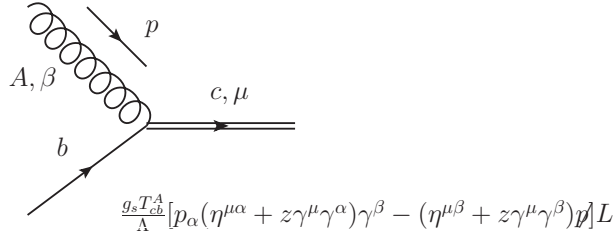}
\caption{Feynman rule used to calculate matrix elements.}
\label{rule}
\end{figure}

Mixing of the spin-1/2 and spin-3/2 states occurs via dimension-five operators of the form:
\begin{equation}
\mathcal{L}_5=i\frac{g}{\Lambda}\bar{\psi}_{\sigma}(\eta^{\sigma\mu}+z\gamma^{\sigma}\gamma^{\mu})\gamma^{\nu}\frac{\lambda^a}{2} P_{R,L}\chi F^a_{\mu\nu}+H.C. ,
\label{mixing} 
\end{equation}where $F^a_{\mu\nu}$ is the field strength tensor of the gauge field. The scale $\Lambda$ sets the strength of the interaction. The corresponding interactions involving a photon or a Z are expected to be subdominant because of the weaker coupling constant. From the Lagrangian terms in Eq.~\ref{mixing} we infer the Feynman rule shown in Fig.~\ref{rule}. The parameter $z$ is called the off-shell parameter as it only affects processes where $t^*_{3/2}$ appears as an intermediate state because of the on-shell condition $\gamma^\mu \psi_\mu=0$. Different phenomenological studies assume different values with $z=-1/4$~\cite{Moussallam:1989nm} and $z=0$~\cite{Kuhn:1984rj} being popular choices. We leave $z$ as a free parameter and will comment on its influence on the $t\bar{t}$ cross section below.

\begin{figure}[h]
\centering
\includegraphics[scale=0.6]{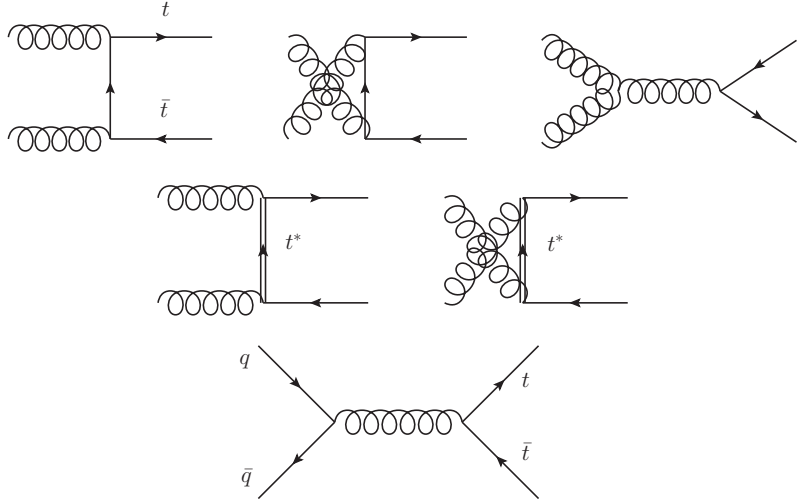}
\caption{Feynman diagrams contributing to $t\bar{t}$ production.}
\label{diagrams}
\end{figure}

The diagrams relevant for $t\bar{t}$ pair production are shown in Fig.~\ref{diagrams}. For gluon-gluon scattering the first three constitute the SM set and we get two additional diagrams from the mixing of $t$ and $t^*_{3/2}$. The final diagram shows the SM production through light quark-antiquark annihilation. At the LHC the dominant contribution to the cross section comes from gluon scattering while at the Tevatron quark-antiquark annihilation dominates. The first deviation from the SM on increasing the top pair energy is expected to come from the interference of the NP and SM diagrams. We note that in contrast with four-particle contact interactions, where the sign of the interference is not fixed, here the vertex of Fig.~\ref{rule} always appears twice in the diagrams of Fig.~\ref{diagrams} leaving no ambiguity in terms of the sign of the interference.

For the sum of the two new diagrams the matrix element is given by: $M_{BSM}^{\mu}\epsilon_{1\mu}$, with $\epsilon^{\mu}_1$ the polarisation vector of one of the initial state gluons. It is possible to show that the matrix element satisfies: $M^{\mu}_{BSM}p_{1\mu}=0$. In fact this applies to each diagram individually and can be explained by the form of the vertex shown in Fig.~\ref{rule}. Therefore for $|M_{BSM}|^2$ it is sufficient to sum covariantly over the gluon helicities. On the contrary, the SM amplitude is not transverse i.e. $M_{SM}^{\mu}p_{1\mu}\neq 0$, which means that summing covariantly over gluon polarisations using $-g_{\mu\nu}$ will introduce spurious contributions. Therefore we need to sum over only the physical polarisations. The same argument was made in \cite{Georgi:1978kx} for $c\bar{c}$ production.

To calculate the matrix element squared we sum over the physical polarisations of the external gluons using an axial gauge projector:  
\begin{eqnarray}
\sum_T \epsilon_{T}^{\mu*}(k)\epsilon_{T}^{\nu}(k)=-g^{\mu\nu}+\frac{k^{\mu}n^{\nu}+k^{\nu}n^{\mu}}{nk}.
\label{proj}
\end{eqnarray} An alternative method is to include diagrams with external ghosts. We calculate the matrix elements squared using FORM~\cite{Vermaseren:2000nd} with the axial gauge type polarisation sum of Eq.~\ref{proj}. Even though contributions from individual diagrams have $n$-dependent terms, the sum is independent of $n$ as expected. Our results were checked using CalcHep~\cite{Pukhov:2004ca} which allows the implementation of a spin-3/2 fermion and employs the ghost diagram method. In this case the only additional diagram is the one where in the third diagram of Fig.~\ref{diagrams} we replace the external gluons with ghosts. 

Summarising the calculation, we have the Feynman rules extracted from the Lagrangian and checked by comparing with the output of LanHEP~\cite{Semenov:2010qt}, the matrix elements squared calculated in FORM and checked by implementing the additional particles and the corresponding vertices in CalcHEP. These were then used to obtain the partonic cross sections by integrating over the phase space using Mathematica. To obtain the hadronic results we integrate over PDFs with our own VEGAS integrator routine. The PDF set used is MSTW2008LO~\cite{Martin:2009iq} and the factorisation and renormalisation scale is set to $m_t=171.3$~GeV\cite{Amsler:2008zzb}. We also note that  higher order calculations exist for the SM $t\bar{t}$ cross section at NLO \cite{Nason:1987xz,Beenakker:1988bq} and several approaches towards NNLO have been taken, see for example \cite{Kidonakis:2008mu,Cacciari:2008zb,Langenfeld:2009wd,Kidonakis:2010dk} and the review in \cite{Kidonakis:2011ca}. However for consistency with the absence of NLO predictions for the BSM contribution in the rest of this paper we use the LO prediction for the SM $t\bar{t}$ cross section.  

The value of $\Lambda$ we use for the calculations is 7$M_{3/2}$. In \cite{Hassanain:2009at} it is calculated that unitarity breaks down for $\sqrt{s}>7M_{3/2}$ by considering gluon mediated scattering in the s-wave. We checked that this bound is reasonable by also calculating the point where unitarity breaks down in $t_{3/2}^*t^*_{3/2}$ scattering using a partial wave amplitude method. The authors in \cite{Hassanain:2009at} suggest using $7M_{3/2}$ as a cut-off for the theory and at the same time setting $\Lambda=7M_{3/2}$. The cross-section sensitivity to this choice will be discussed later in this Section. 

\begin{figure}[h]
\centering
\includegraphics[scale=0.7]{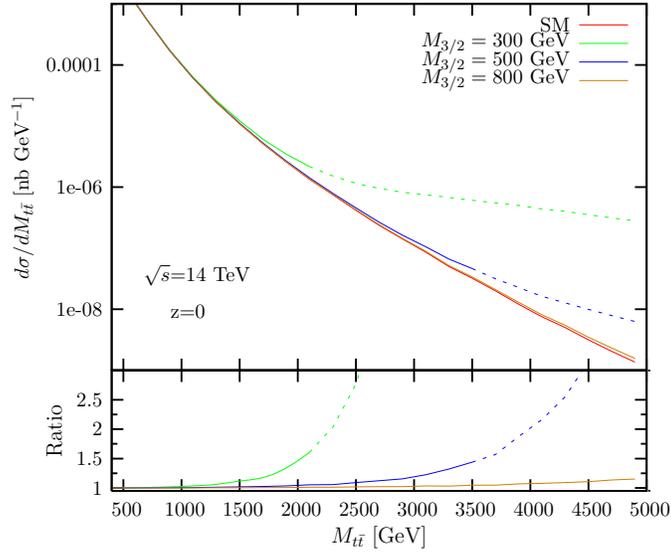}
\caption{Differential cross section and ratio to SM prediction for 14 TeV.}
\label{ttbar}
\end{figure}

\begin{figure}[h]
\centering
\includegraphics[scale=0.7]{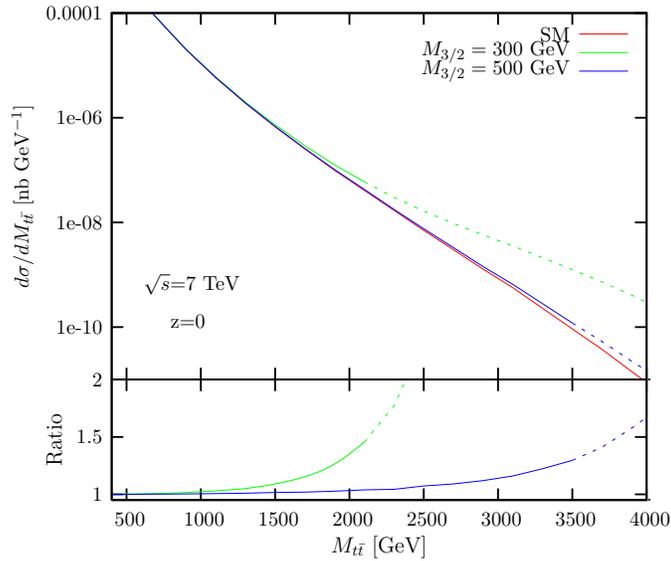}
\caption{Differential cross section and ratio to SM prediction for 7~TeV.}
\label{ttbar7}
\end{figure}

We investigate how the cross section varies by adjusting the free parameters in the model. Firstly the $t\bar{t}$ cross section at the LHC is studied as a function of the mass of $t_{3/2}^*$. The differential cross section $d\sigma/dM_{t\bar{t}}$ at the LHC at 14 TeV is shown in Fig.~\ref{ttbar} for three different masses of $t_{3/2}^*$ with $z=0$. On the same plot we also show the ratio to the SM prediction. The results show an excess in the cross section at large $t\bar{t}$ masses. The deviation gets rapidly less important with increasing $t_{3/2}^*$ mass. This is expected as the mass of the new state suppresses the matrix element both through the propagator and especially through the scale $\Lambda$ which we set proportional to the mass.
The corresponding results for the LHC at 7 TeV are shown in Fig.~\ref{ttbar7}, where we see that the reach mass region is rapidly narrowed as we are more limited by energy. For 7 TeV we omit the results for $M_{3/2}=800$ GeV as the deviation from the SM is not visible in the graph. In the plots of Fig.~\ref{ttbar} and \ref{ttbar7} we change the lines from solid to dashed at the point where the partonic CoM energy reaches the proposed cut-off of $7M_{3/2}$. Imposing this cut-off significantly decreases the chance of calculating a measurable deviation from the SM. For example, Fig.~\ref{ttbar} shows that for $M_{3/2}=300$ GeV the cut-off will be 2.1~TeV which implies that our description becomes unreliable before we reach the region where the excess becomes detectable. 
This cut-off introduced to prevent unitarity violation, contrary to what we expect from the underlying theory, behaves as a step function. Unitarity violation can also be avoided by the construction of appropriate form factors which are then used to damp the growth of the cross sections at high energies. This is discussed in \cite{Moussallam:1989nm} in connection with spin-3/2 excitations. The introduction of form factors is a common technique used to avoid unitarity violation in BSM models, for example they are introduced for anomalous triple gauge boson couplings in \cite{Baur:1989gk} and adopted by the corresponding experimental searches. However the form of the form factor is not universal. In the absence of knowledge of the precise form of the form factor and the fact that this will introduce an additional uncertainty in the calculation, we choose not to introduce a form factor here. 

\begin{figure}[h]
\begin{minipage}[b]{0.5\linewidth}
\centering
\includegraphics[trim=2.1cm 0 0 0,scale=0.6]{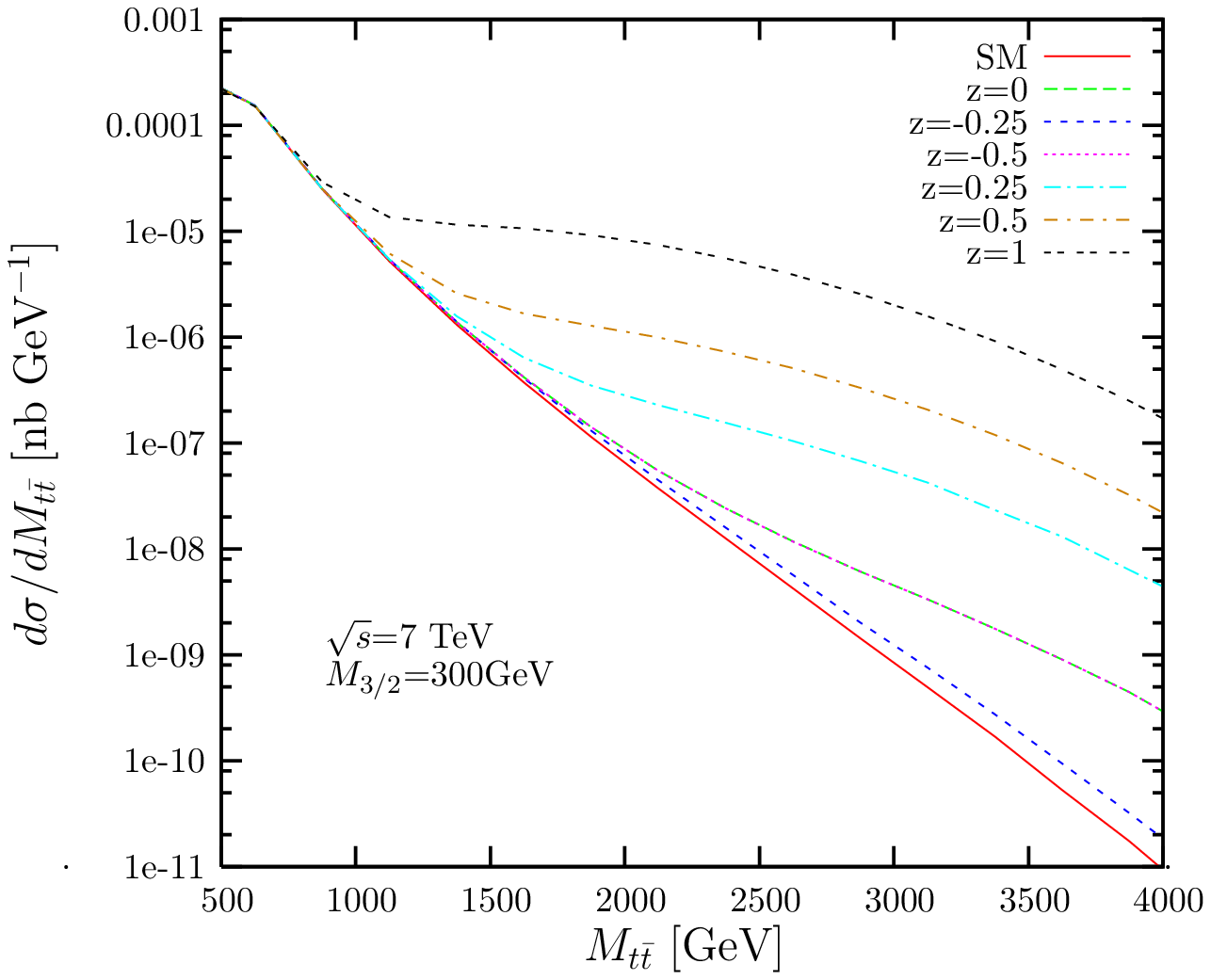}
\caption{Differential cross section for $M_{3/2}$=300 GeV for different values of $z$ at 7 TeV.}
\label{ttbar7z}
\end{minipage}
\hspace{0.5cm}
\begin{minipage}[b]{0.5\linewidth}
 \centering
\includegraphics[trim=2.9cm 0 0 0,scale=0.6]{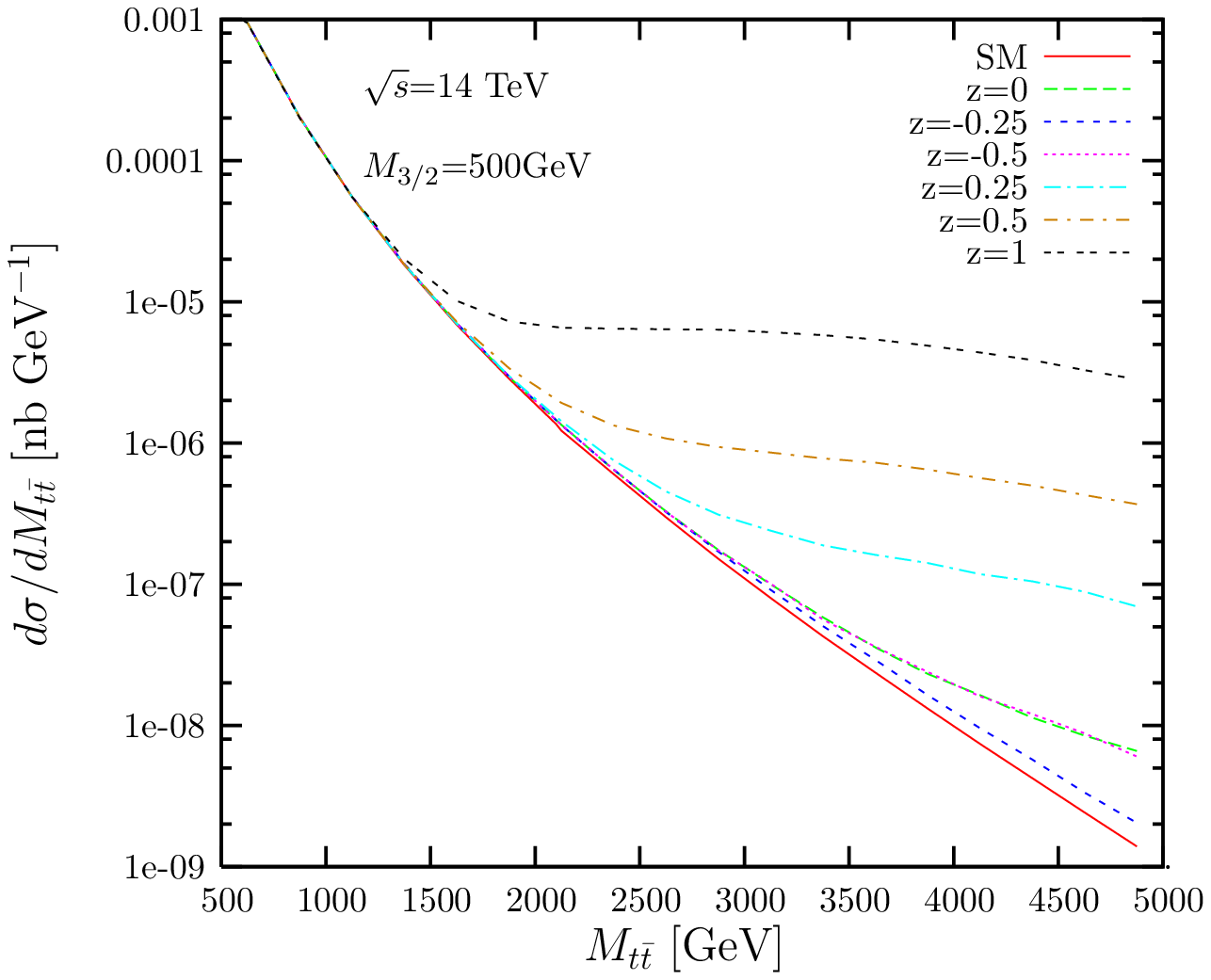}
\caption{Differential cross section for $M_{3/2}$=500 GeV for different values of $z$ at 14 TeV.}
\label{zeff}
\end{minipage}
\end{figure}

In addition to the mass of the excited top, the cross section depends on the off-shell parameter $z$. A solid value of $z$ has not been established, with different arguments existing in the literature. Possible values for $z$ in the context of spin-3/2 baryon resonances which are described by the same Lagrangian are discussed in \cite{Benmerrouche:1989uc}. The importance of $z$ for the top pair mass differential distribution is shown in Fig.~\ref{ttbar7z} for $M_{3/2}=$300 GeV at 7~TeV and Fig.~\ref{zeff} for $M_{3/2}=$500 GeV at 14~TeV. We notice that the value of $z$ significantly affects the results. The BSM matrix element squared is a complicated function of $z$ with terms up to $z^4$. The coefficients of the different powers of $z$ depend on the mass of the excited top and the centre-of-mass energy and generally increase with power. The value of $z$ effectively acts as another coupling strength controlling the cross section. The minimum contribution to the cross section occurs at $z=-0.25$ and then increases symmetrically around this minimum i.e. with the result for $z=-0.5$ being very close to that of $z=0$, with the corresponding lines coinciding in the plots. This applies in the region where the dominant contribution comes from the pure BSM contribution which is a 4th order polynomial in $z$. The interference contribution is a second order polynomial but it rapidly becomes subdominant as the top pair invariant mass increases. The region where a significant deviation from the SM occurs is completely dominated by the pure BSM contribution. 

\begin{figure}[h]
\begin{minipage}[b]{0.5\linewidth}
\centering
\includegraphics[trim=2.6cm 0 0 0 ,scale=0.6]{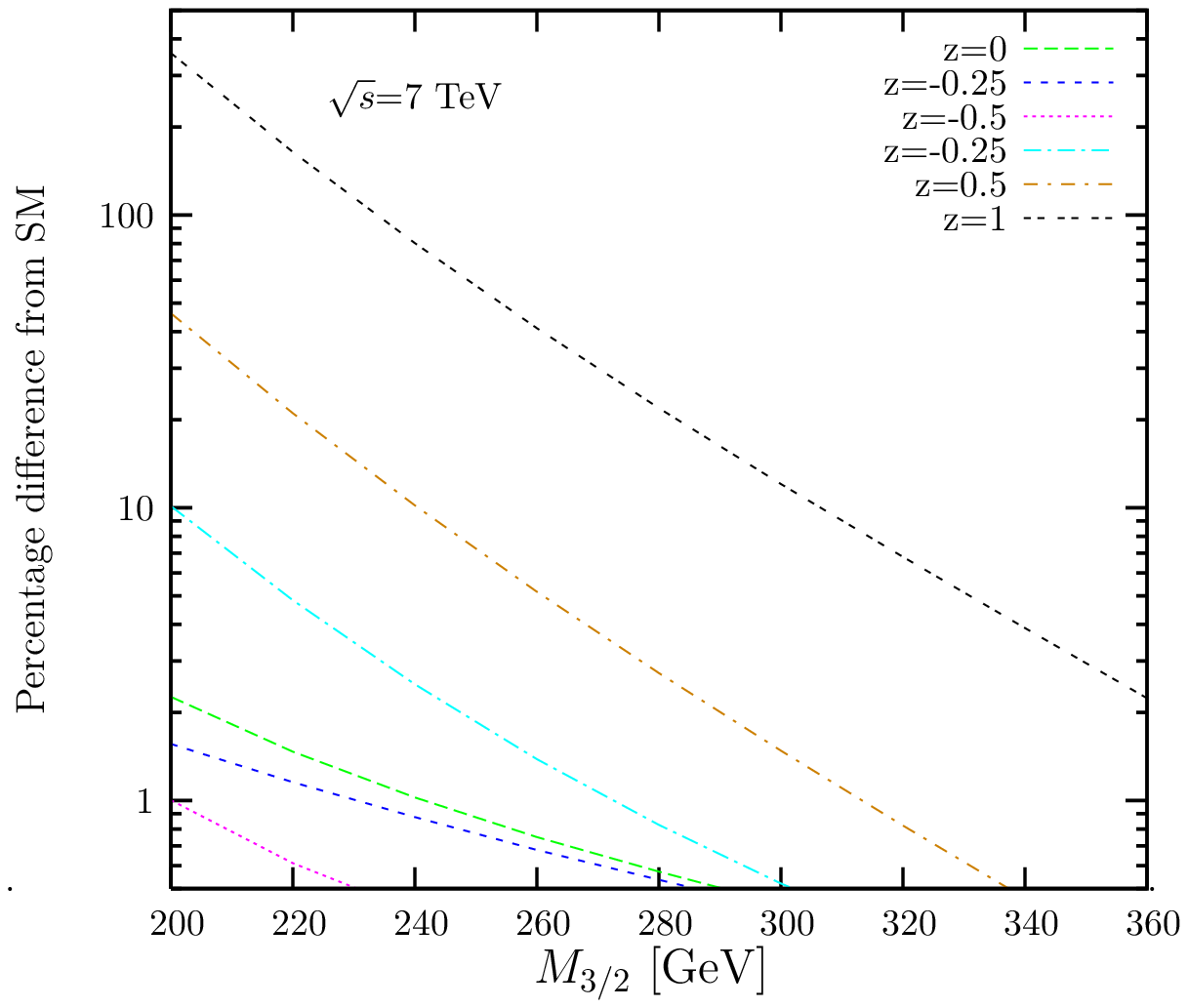}
\caption{Percentage difference from the SM prediction for the total $t\bar{t}$ production cross section at 7~TeV.}
\label{7total}
\end{minipage}
\hspace{0.5cm}
\begin{minipage}[b]{0.5\linewidth}
\centering
\includegraphics[trim=3.5cm 0 0 0 ,scale=0.6]{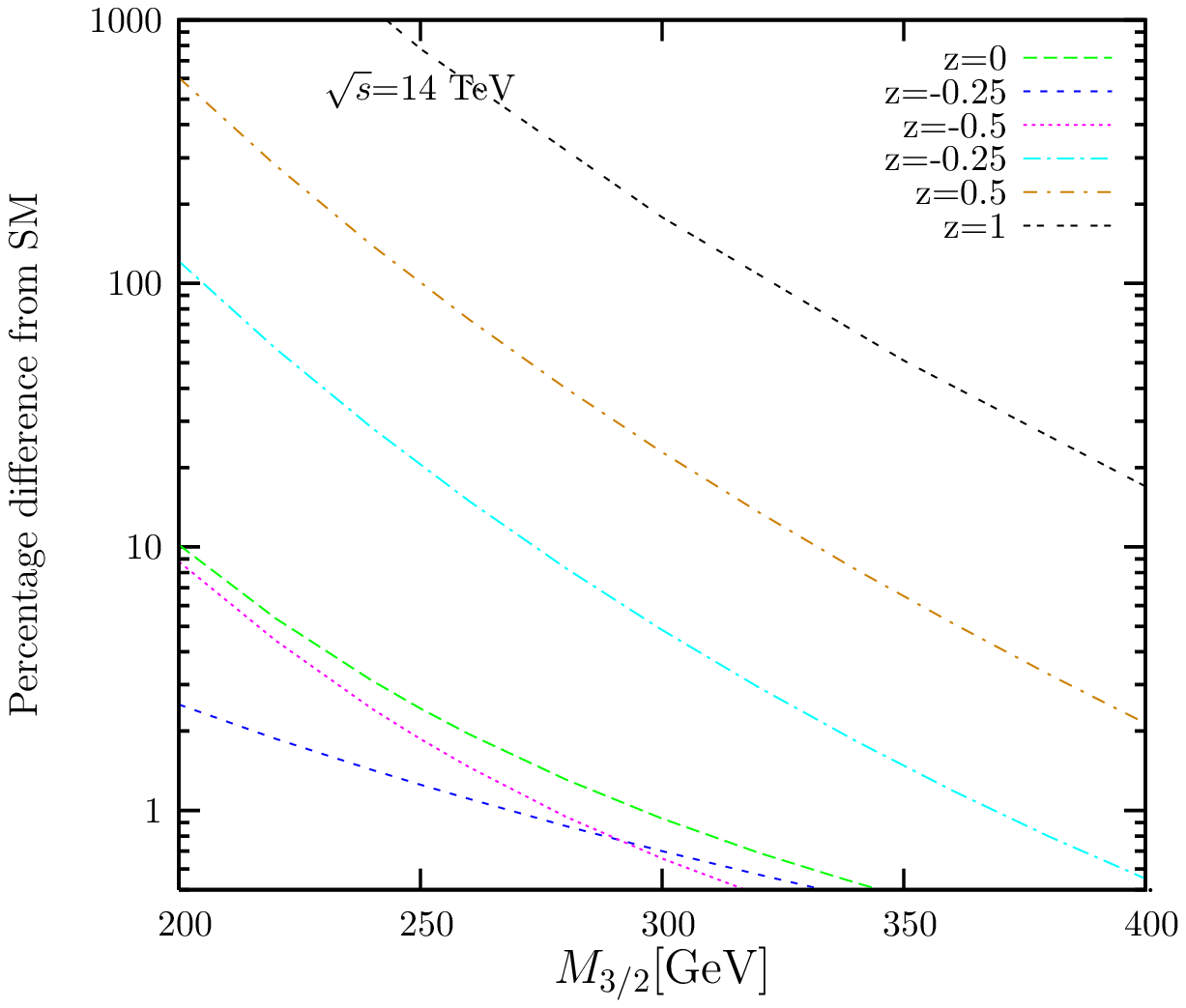}
\caption{Percentage difference from the SM prediction for the total $t\bar{t}$ production cross section at 14~TeV.}
\label{total}
\end{minipage}
\end{figure}

We also consider the total $t\bar{t}$ cross section at the LHC as a function of the mass of $t_{3/2}^*$.  The deviation from the SM falls quickly to the percent level for the most popular choices of negative and zero $z$, as shown in Fig.~\ref{7total} and \ref{total} for 7 and 14~TeV respectively. Therefore the most promising way to detect any deviation from the SM is to look for an excess of events of high $t\bar{t}$ mass. We also note a subtle difference between the two sets of graphs: Figs.~\ref{ttbar7z},\ref{zeff} and \ref{7total},\ref{total}. As noted above, in the first set the effect of the $z$ choice which we see in the graph at high top pair invariant masses comes from the pure BSM contribution. The contribution of the interference term, which can be both negative and positive depending on the CoM energy and the value of $z$, is rapidly overtaken by the pure signal with increasing top pair invariant mass. On the other hand when we consider the deviation of the total cross section from the SM prediction then the interference contribution, which dominates over the pure signal in the low $t\bar{t}$ mass region, becomes important. This is related to the fact that the cross section is a rapidly falling function of the top pair invariant mass. This explains what would appear to be a discrepancy for example between the results at $M_{3/2}=$300~GeV for $z=0$ and $z=-0.5$ in Fig.~\ref{ttbar7z} where the two lines (pink and green) coincide and the corresponding results in Fig.~\ref{7total} where the $z=0$ line appears significantly higher. 

\begin{figure}[h]
\begin{minipage}[b]{0.5\linewidth}
\centering
\labellist
\footnotesize\hair 2pt
\pinlabel \rotatebox{90}{$M_{3/2}$} at 140 486
\pinlabel \rotatebox{90}{$\text{[GeV]}$} at 140 520
\endlabellist

\includegraphics[scale=0.75]{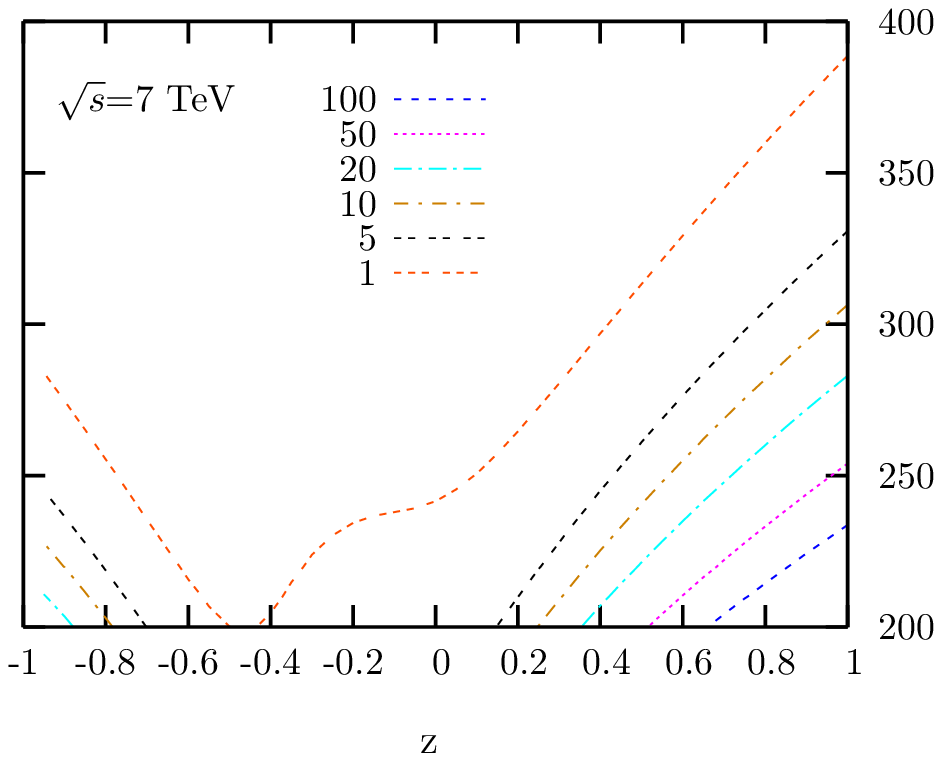}
\caption{Contour plot of percentage difference from the SM prediction for the total $t\bar{t}$ cross section at 7~TeV.}
\label{contour7}
\end{minipage}
\hspace{0.5cm}
\begin{minipage}[b]{0.5\linewidth}
\labellist
\footnotesize\hair 2pt
\pinlabel \rotatebox{90}{$M_{3/2}$} at 140 486
\pinlabel \rotatebox{90}{$\text{[GeV]}$} at 140 520
\endlabellist
\includegraphics[scale=0.75]{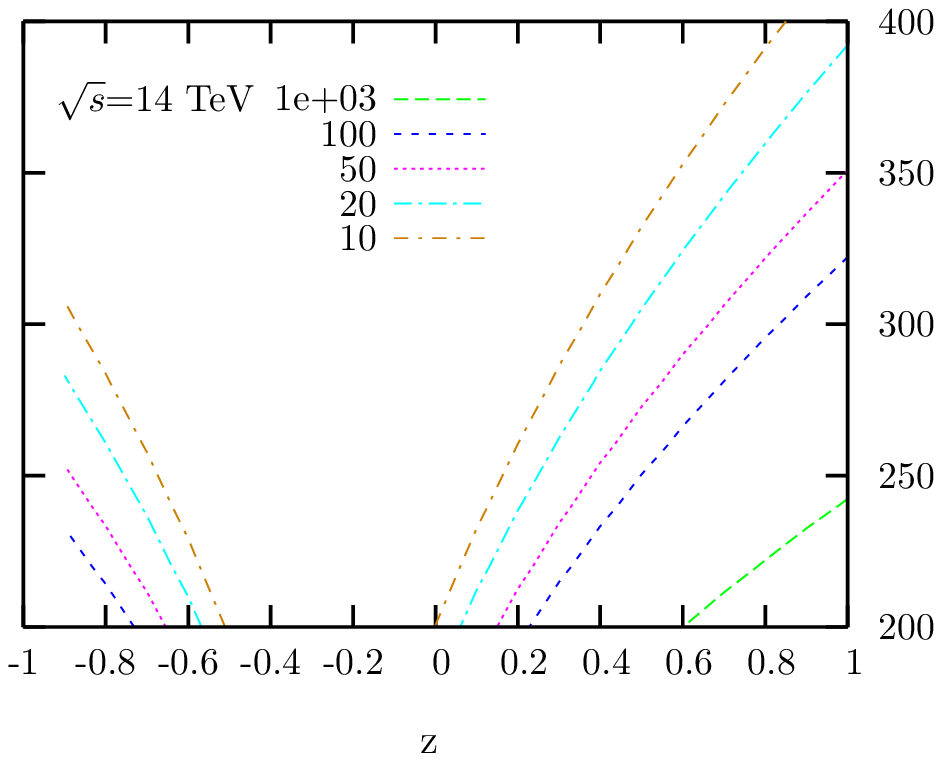}
\caption{Contour plot of percentage difference from the SM prediction for the total $t\bar{t}$ cross section at 14~TeV.}
\label{contour} 
\end{minipage}
\end{figure}

\begin{figure}[h]
\begin{minipage}[b]{0.5\linewidth}
\labellist
\footnotesize\hair 2pt
\pinlabel \rotatebox{90}{$M_{3/2}$} at 140 486
\pinlabel \rotatebox{90}{$\text{[GeV]}$} at 140 520
\endlabellist
\centering
\includegraphics[scale=0.75]{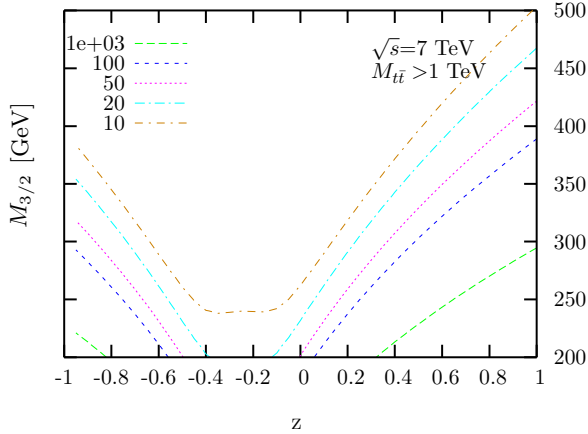}
\end{minipage}
\hspace{0.5cm}
\begin{minipage}[b]{0.5\linewidth}
\centering
\labellist
\footnotesize\hair 2pt
\pinlabel \rotatebox{90}{$M_{3/2}$} at 140 486
\pinlabel \rotatebox{90}{$\text{[GeV]}$} at 140 520
\endlabellist
\includegraphics[scale=0.75]{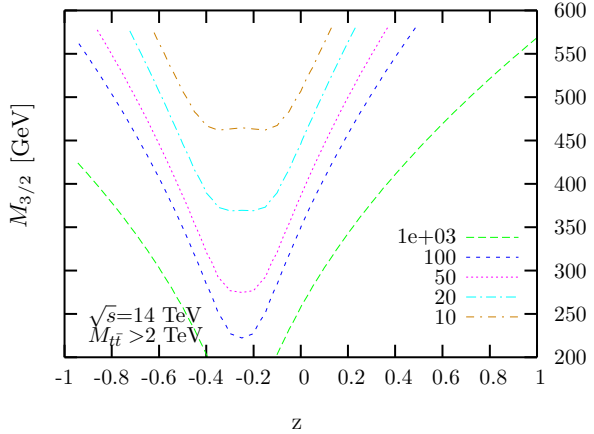}
\end{minipage}
\caption{Percentage difference from the SM prediction for the total $t\bar{t}$ cross section at 7~TeV with a $t\bar{t}$ invariant mass cut of 1~TeV and 14~TeV with a cut of 2~TeV.}
\label{714cut}
\end{figure}

To simultaneously study the dependence on both parameters we calculate contour plots of the percentage difference from the SM prediction as a function of the $t^*_{3/2}$ mass and the off-shell parameter $z$ at 7 and 14 TeV in Figs.~\ref{contour7} and \ref{contour} respectively. We note that the range of mass and $z$ values for which the deviation is significant is rather limited. Only a very narrow window of low mass and high $z$  will give a deviation larger than the uncertainties in the SM prediction. In order to make the search more sensitive to the NP contribution we need to impose a cut on the invariant mass of the top pair as otherwise the deviation will fall within the uncertainty of the SM prediction. The results for cuts of 1~TeV and 2~TeV for 7~TeV and 14~TeV respectively are shown in Fig.~\ref{714cut}. These plots can be used as a guide for the reach of the LHC and the exclusion region that can be set given both the theoretical and experimental uncertainties in the measurement of the cross section. Of course we need to stress that imposing a cut on the invariant mass poses two issues, the first one being that we depend more on the energy region above the proposed cut-off where the effective theory description becomes less reliable. The second problem is that the experimental errors are expected to increase in the high mass region. Nevertheless we see that a narrow region of parameter phase space can be easily excluded even at 7~TeV with 14~TeV being more promising in terms of the extent of the exclusion region.

\begin{figure}[h]
\centering
\labellist
\footnotesize\hair 2pt
\pinlabel \rotatebox{0}{$\text{[GeV]}$} at 490 496
\endlabellist
\includegraphics[scale=0.8]{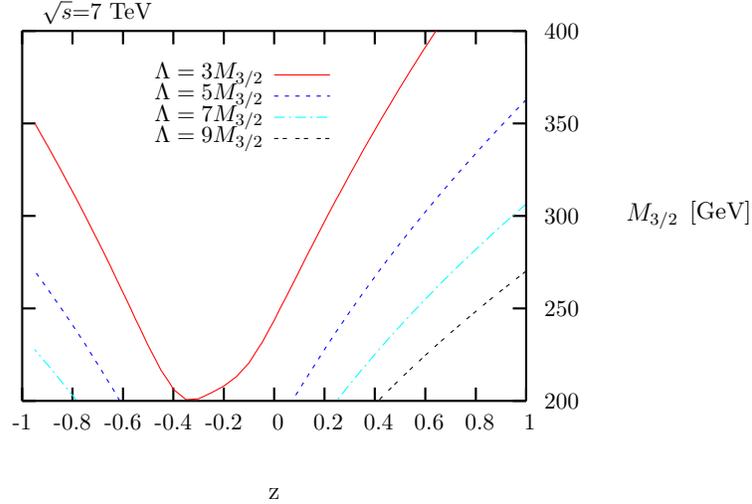}
\caption{Contour plots for a 10\% deviation from the SM prediction for the total $t\bar{t}$ cross section for different values of $\Lambda$ for 7~TeV.}
\label{contourl}
\end{figure}

For the plots above we take $\Lambda=7M_{3/2}$. However this is only an estimate coming from a unitarity argument. $\Lambda$ is effectively another free  parameter which is expected to be of the order of the mass of the excited top but exact numerics are unknowable. To investigate the impact of $\Lambda$ on the deviation from the SM we show a contour plot for $\Lambda$ set to various multiples of $M_{3/2}$. We only show the 10\% plot for 7 TeV in Fig.~\ref{contourl} as we consider this to be close to the current experimental accuracy of the top pair production measurement. The strong dependence on $\Lambda$ is expected as the matrix element squared for the new diagrams is proportional to $\Lambda^{-2}$ for the interference and $\Lambda^{-4}$ for the signal. This plot serves as a guide for the exclusion region depending on the preferred value of $\Lambda$. Considering the modification of the $t\bar{t}$ cross section will be complimentary to the search for single and pair production of the spin-3/2 top. To date there have been no direct searches for spin-3/2 excited tops, but searches have been performed at the Tevatron for spin-1/2 top excitations. The mass bound set for a fourth generation top quark with SM coupling from pair production is 311 GeV \cite{Pleier:2008ig}. As noted in \cite{Hassanain:2009at} the mass bound based on the Tevatron data for a fourth generation quark is not expected to rise significantly above 300~GeV. In order to translate the $t\bar{t}$ experimental results into a bound on the mass of the excited top a more extended study going beyond the parton level is required involving taking into consideration detector effects, which is beyond the scope of this paper.

\begin{figure}[h]
\begin{minipage}[b]{0.5\linewidth}
\centering
\includegraphics[trim=1.5cm 0 0 0 ,scale=0.6]{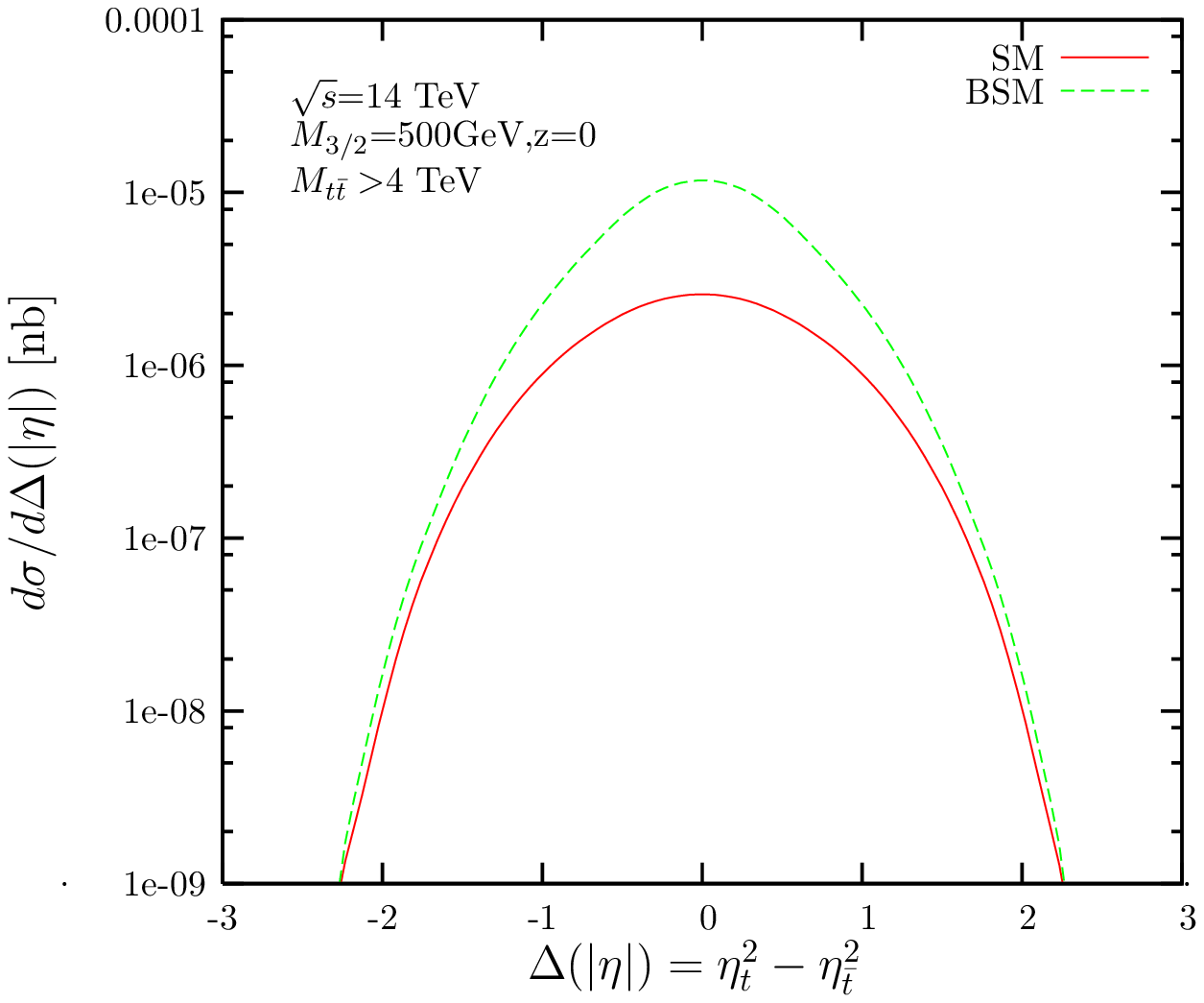}
\end{minipage}
\hspace{0.5cm}
\begin{minipage}[b]{0.5\linewidth}
\centering
\includegraphics[trim=1.9cm 0 0 0 ,scale=0.6]{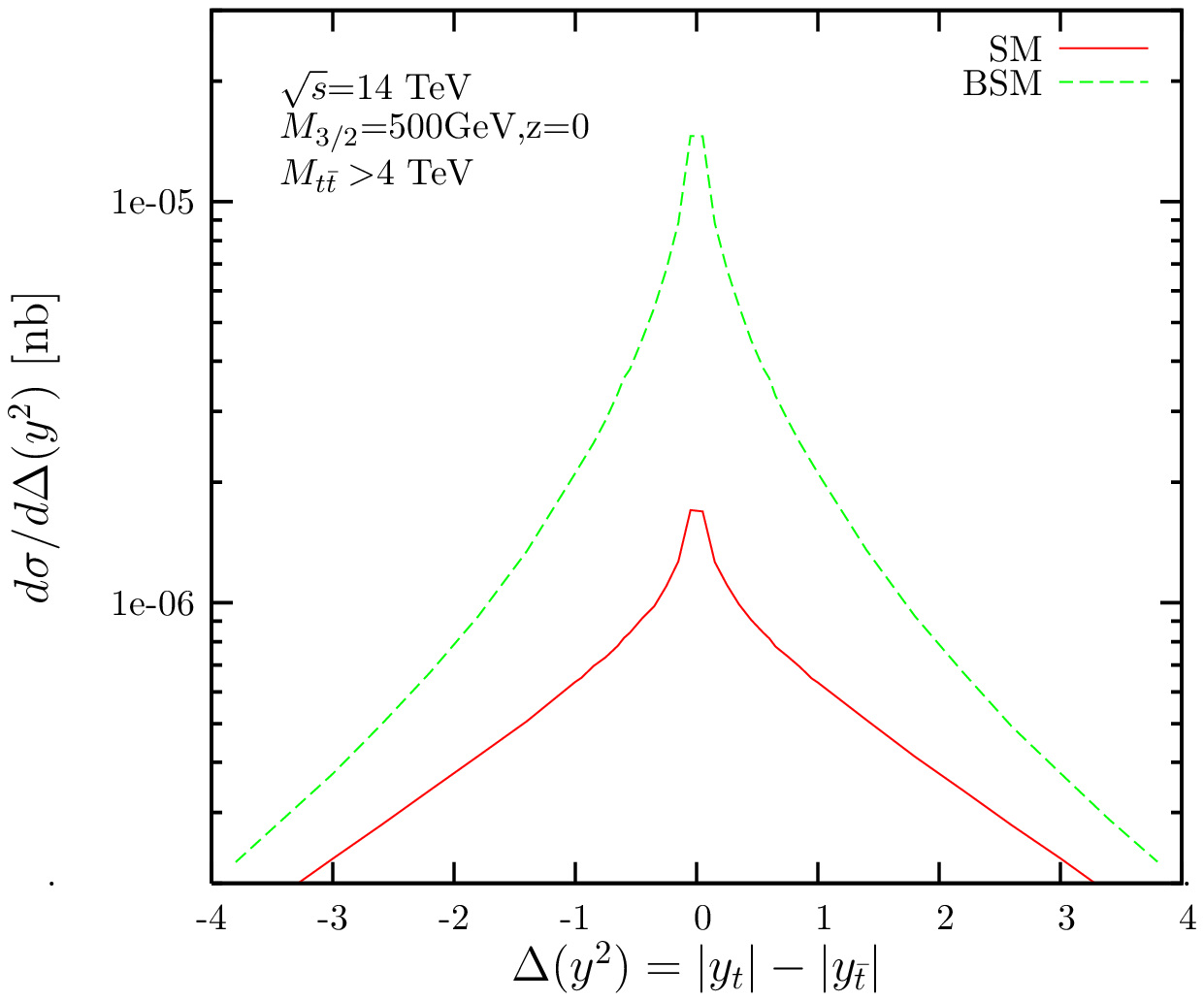}
\end{minipage}
\caption{Differential cross sections for $\Delta(|\eta|)$ and $\Delta(y^2)$.}
\label{deltay}
\end{figure}

Finally we study the angular distribution of the top-antitop pairs by using the variables used in the definition of the forward-backward asymmetry measured by CDF. These are the difference in the absolute pseudorapidities $\Delta(|\eta|)=|\eta_t|-|\eta_{\bar{t}}|$ and the difference of the squares of the rapidities $\Delta(y^2)=y^2_t-y^2_{\bar{t}}$. These variables are also used for searches at the LHC, see for example the CMS study \cite{CMS}. Our signal would not contribute to the asymmetry as shown in the plots of Fig.~\ref{deltay}  for $M_{3/2}=500$ GeV and 14~TeV.

We note that to ensure we get an observable difference in the angular distribution plots we had to impose a high $M_{t\bar{t}}$ cut. The excess above the SM prediction is more significant in the central regions for both  $\Delta(|\eta|)$ and $\Delta(y^2)$. This is related to the presence of a heavy particle in the propagator which does not favour small angle forward scattering. In practice, due to detector restrictions a cut has to be imposed on the rapidity of the top with central tops experimentally preferred. This favours the signal over the background as the excess is more important in the central regions.
\section{Comparison with spin-1/2 excited top}
\begin{figure}[h]
\begin{minipage}[b]{0.5\linewidth}
\centering
\includegraphics[trim=1.9cm 0 0 0,scale=0.6]{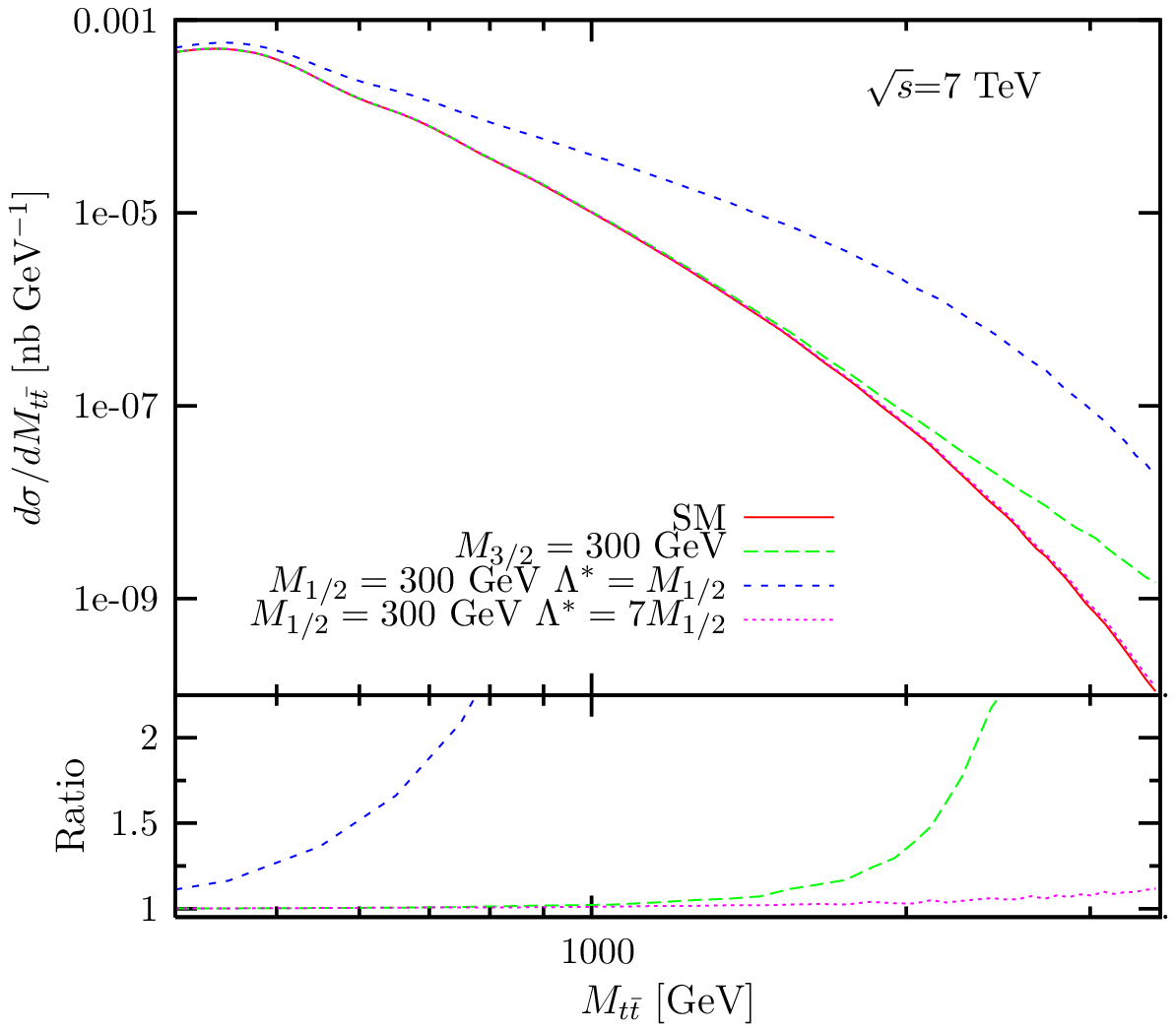}
\end{minipage}
\hspace{0.5cm}
\begin{minipage}[b]{0.5\linewidth}
\centering
\includegraphics[trim=2.9cm 0 0 0,scale=0.6]{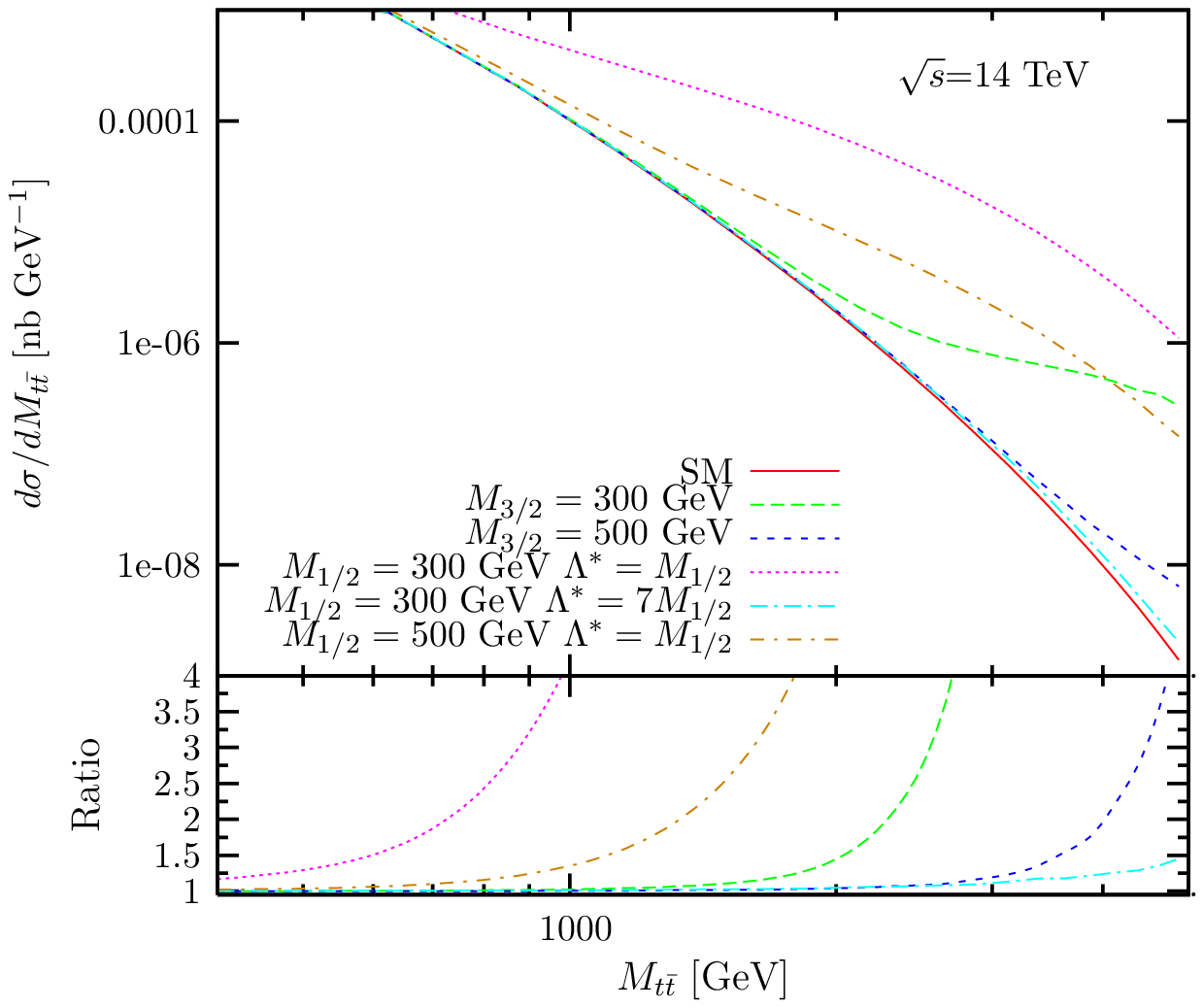}
\end{minipage}
\caption{Differential cross-section for different masses for both spin-1/2 and spin-3/2 excitations and the corresponding ratio to the SM prediction.}
\label{spin12}
\end{figure}

For comparison purposes we also show results for the contribution of a possible spin-1/2 top quark excitation to the $t\bar{t}$ cross section. Excited spin-1/2 quarks and leptons which mix with SM quarks and leptons through the emission of a gauge boson have been extensively studied in the literature. The Lagrangian term we use to extract the Feynman rules is \cite{Baur:1989kv}:
\begin{equation}
\mathcal{L}=\frac{1}{2\Lambda^*}\bar{t}^*_R\sigma^{\mu\nu}g_s\frac{\lambda^a}{2}G^a_{\mu\nu}t_L+H.C.
\label{lspin1/2}
\end{equation}
The scale $\Lambda^*$ is taken in \cite{Baur:1989kv} and other relevant studies to be equal to the mass of the excited quark. The results for the differential cross section and the ratio to the SM prediction are shown in Fig.~\ref{spin12} for the LHC at 7 and 14 TeV. 
We note here that if the scale $\Lambda^*$ is taken to be 7$M_{1/2}$, as the corresponding scale in the spin-3/2 case, the effect of the spin-3/2 excitation is significantly larger for excited tops of the same mass, which is related to the extra momentum factors in the propagator. But if we use $\Lambda^*=M_{1/2}$ and $\Lambda=7M_{3/2}$ then the spin-3/2 contribution is strongly suppressed by the inverse powers of $\Lambda$ and the spin-1/2 effect is prevailing. Another observation from the graph (see green and pink lines) is the difference in the shapes of the two contributions. In the case of spin-3/2 tops the cross section rises more rapidly above the SM prediction suffering more strongly from unitarity violation. The results for a spin-1/2 excited top will be further used in the following Section where we compare with a dimension-six operator model.
 
\section{Comparison with other BSM operators affecting $t\bar{t}$ production}
\begin{figure}[h]
\centering
\includegraphics[scale=0.6]{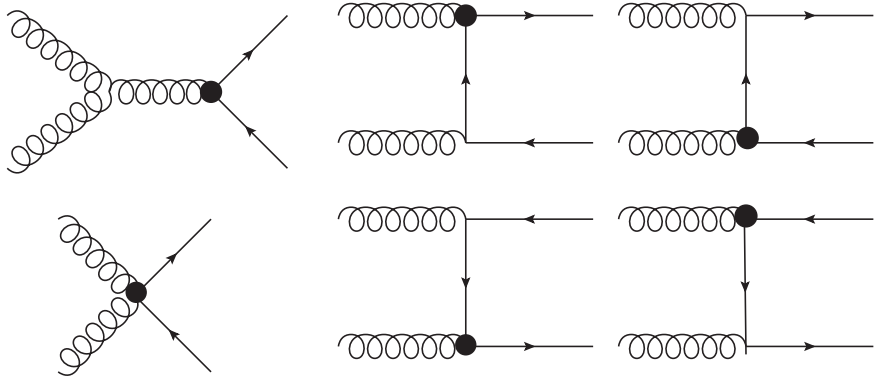}
\caption{Feynman diagrams from the chromomagnetic operator.}
\label{dim6diag}
\end{figure}

In this section we compare the $t\bar{t}$ signal from the spin-3/2 excited top to other BSM scenarios. In \cite{Degrande:2010kt} the authors consider a set of BSM operators of dimension-six. The long list of possible operators \cite{Buchmuller:1985jz} is reduced by considering the underlying symmetries and the equations of motion. A set of operators which affect top quark pair production is collected. This set includes the chromomagnetic dipole moment of the top quark and four-fermion operators. Here for the comparison we pick the chromomagnetic operator: 
\begin{equation}
\mathcal{O}_{hg}=[(H\bar{Q})\sigma^{\mu\nu}T^At]G_{\mu\nu}^A.
\end{equation}The coefficient of this operator in the Lagrangian is given by $c_{hg}/\Lambda^2$, with $c_{hg}$ a dimensionless constant of order one and $\Lambda$ a TeV scale. The operator can be related to theories of top compositeness and has been studied extensively in the literature, see for example \cite{Atwood:1994vm,Cheung:1995nt}, and more recently \cite{Hioki:2009hm,Degrande:2010kt} and references therein. It modifies the $gt\bar{t}$ coupling and its effect is expected to dominate over four-fermion operators at the LHC where 85\% of $t\bar{t}$ production comes from gluon-gluon fusion. The additional Feynman diagrams for $gg\rightarrow t\bar{t}$ originating
 from this operator are shown in Fig.~\ref{dim6diag}. 
For the process $q\bar{q}\rightarrow t\bar{t}$ the diagram is identical to the SM one with the $gt\bar{t}$ vertex replaced by a blob. Each vertex denoted by a blob is suppressed by two powers of $\Lambda$. Therefore the leading effects of the operator are expected to come from the interference of these diagrams with the three diagrams of the SM. We also note that $c_{hg}$ can be positive or negative and therefore either increase or decrease the cross section in the case where only the interference is taken into account. In \cite{Degrande:2010kt} it is argued that the contribution of $\Lambda^{-4}$ terms can be neglected for sufficiently low energies and only the interference of the dimension-six operator with the SM is taken into consideration. We reproduced the results of \cite{Degrande:2010kt,Hioki:2009hm} for the matrix elements squared and the total partonic cross section using FORM. We then used these results to obtain the distributions for different kinematical variables to compare with our results for the spin-1/2 and spin-3/2 excited tops. 

As a starting point and a check we summarise our results for the SM cross section for the LHC and the Tevatron. The SM results are summarised in Table 1. These are tree level results. We note that \cite{Degrande:2010kt} uses CTEQ6L1 PDFs. We compare our results for MSTW2008LO and CTEQ6L1 for both the SM in Table \ref{smtable} and the additional $\mathcal{O}_{hg}$ contribution in Table \ref{bsmtable}.
\begin{table} 
\caption{Comparison of results for the SM in pb}
\begin{center}
    \begin{tabular}{ | c | c | c | c |}
    \hline
    Set  & SM LHC7 & SM LHC14 &SM Tevatron   \\ \hline
    MSTW2008LO & 128 & 716 & 7.04  \\ \hline
    CTEQ6L1 & 107 & 632 & 6.63\\ \hline
   \end{tabular}
\end{center}
\label{smtable}
\end{table}

\begin{table} 
\caption{Comparison of results for $\mathcal{O}_{hg}$ in pb}
\begin{center}
    \begin{tabular}{ | c | c | c | c |}
    \hline
    Set  & $\mathcal{O}_{hg}$ LHC7 & $\mathcal{O}_{hg}$ LHC14 & $\mathcal{O}_{hg}$ Tevatron   \\ \hline
    MSTW2008LO & 32 & 180 & 1.56  \\ \hline
    CTEQ6L1 & 28 & 165 & 1.51\\ \hline
   \end{tabular}
\end{center}
\label{bsmtable}
\end{table}
The table values for $\mathcal{O}_{hg}$ show the increase in the cross section for $c_{hg}/\Lambda^2$=1~TeV$^{-2}$. For other values of $c_{hg}/\Lambda^2$ one can appropriately rescale as this is the contribution of the interference only. 
In the calculation we set the renormalisation and factorisation scales equal to the top mass. The difference between the two PDF sets is explained by the difference in the value of $\alpha_s$ and the difference in the gluon PDF. For MSTW2008LO $\alpha_s(m_t)=0.125$ while for CTEQ6L1 $\alpha_s(m_t)=0.118$ which leads to a difference of more than 10\% between the two sets for the SM cross section which is proportional to $\alpha_s^2$. At the LHC at 14~TeV this almost completely accounts for the difference. At the Tevatron and the LHC at 7~TeV the difference is a compination of the $\alpha_s$ value and the PDF values. At the LHC at 7~TeV it is still dominated by $\alpha_s$ while if we decompose the cross section into contributions from gluon-gluon and quark-antiquark annihilation we can further infer that at the Tevatron the difference is a more complicated function of $\alpha_s$ and the gluon and quark distributions at high $x$. Similar considerations apply for the NP contribution. 

\begin{figure}[h]
\begin{minipage}[b]{0.5\linewidth}
\centering
\includegraphics[trim=1.5cm 0 0 0 ,scale=0.6]{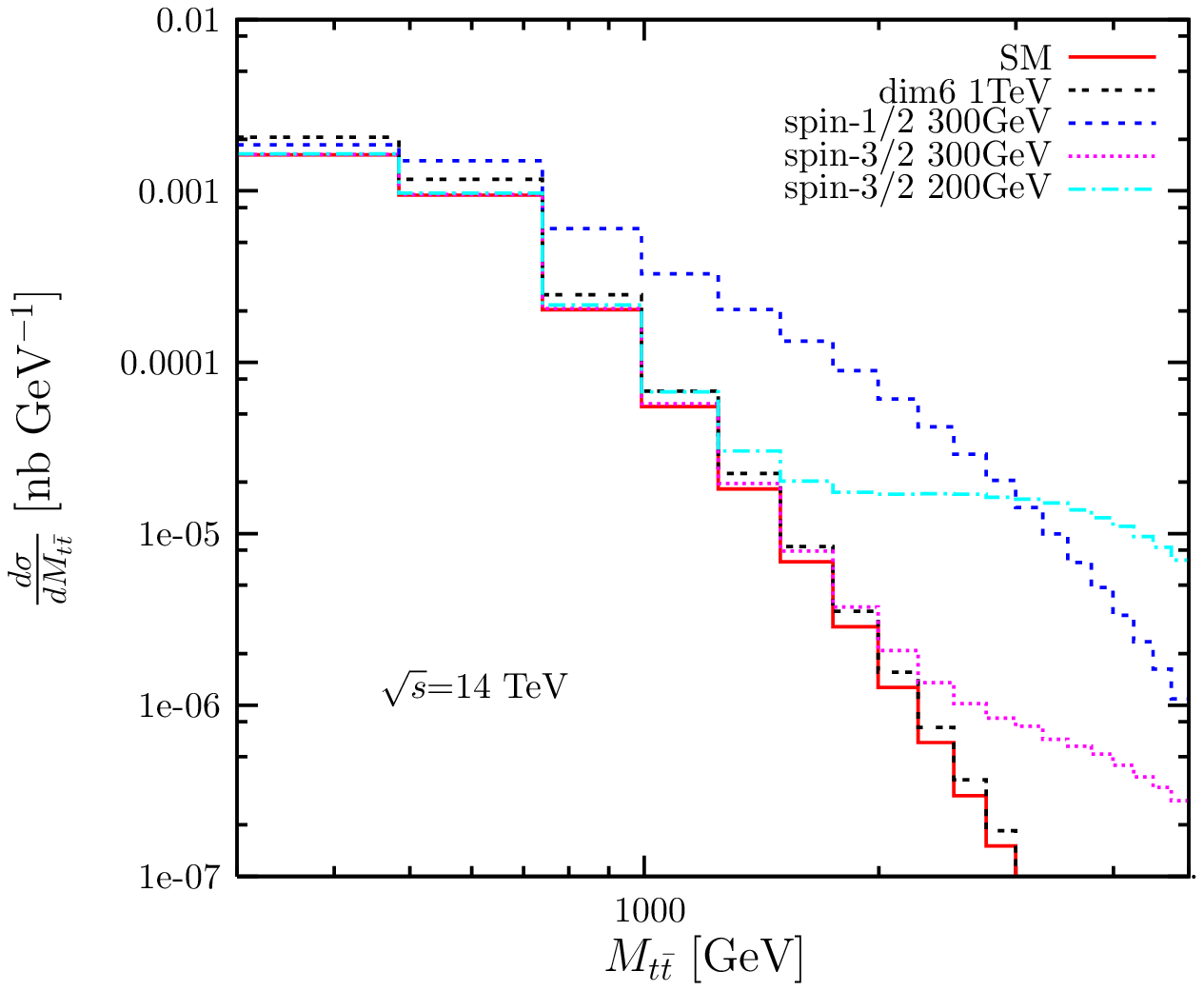}
\end{minipage}
\hspace{0.5cm}
\begin{minipage}[b]{0.5\linewidth}
\centering
\includegraphics[trim=1.5cm 0 0 0 ,scale=0.6]{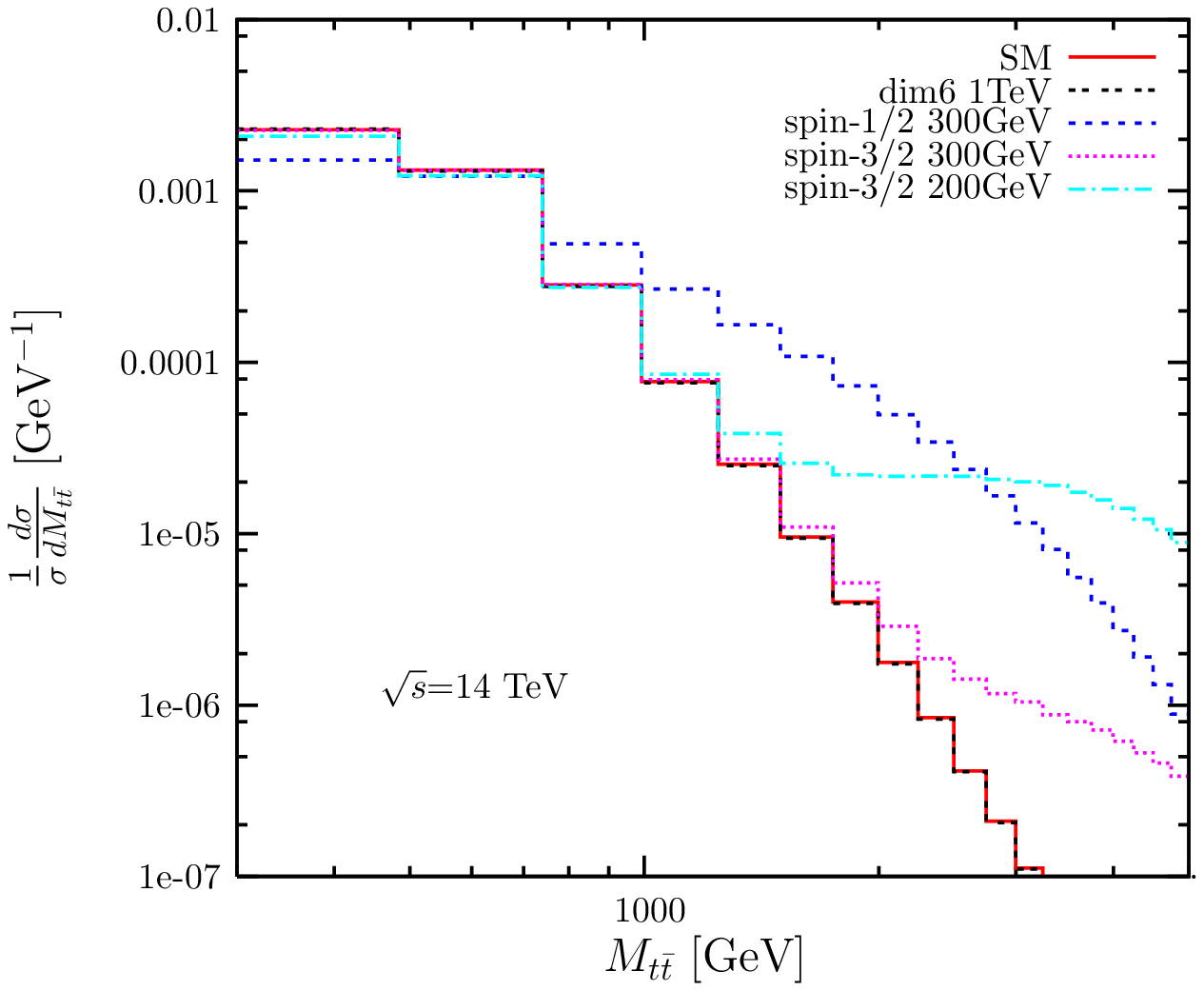}
\end{minipage}
\caption{Unnormalised and normalised differential cross section for $M_{t\bar{t}}$.}
\label{mttunno2}
\end{figure}

\begin{figure}[h]
\begin{minipage}[b]{0.5\linewidth}
\centering
\includegraphics[trim=1.5cm 0 0 0 ,scale=0.6]{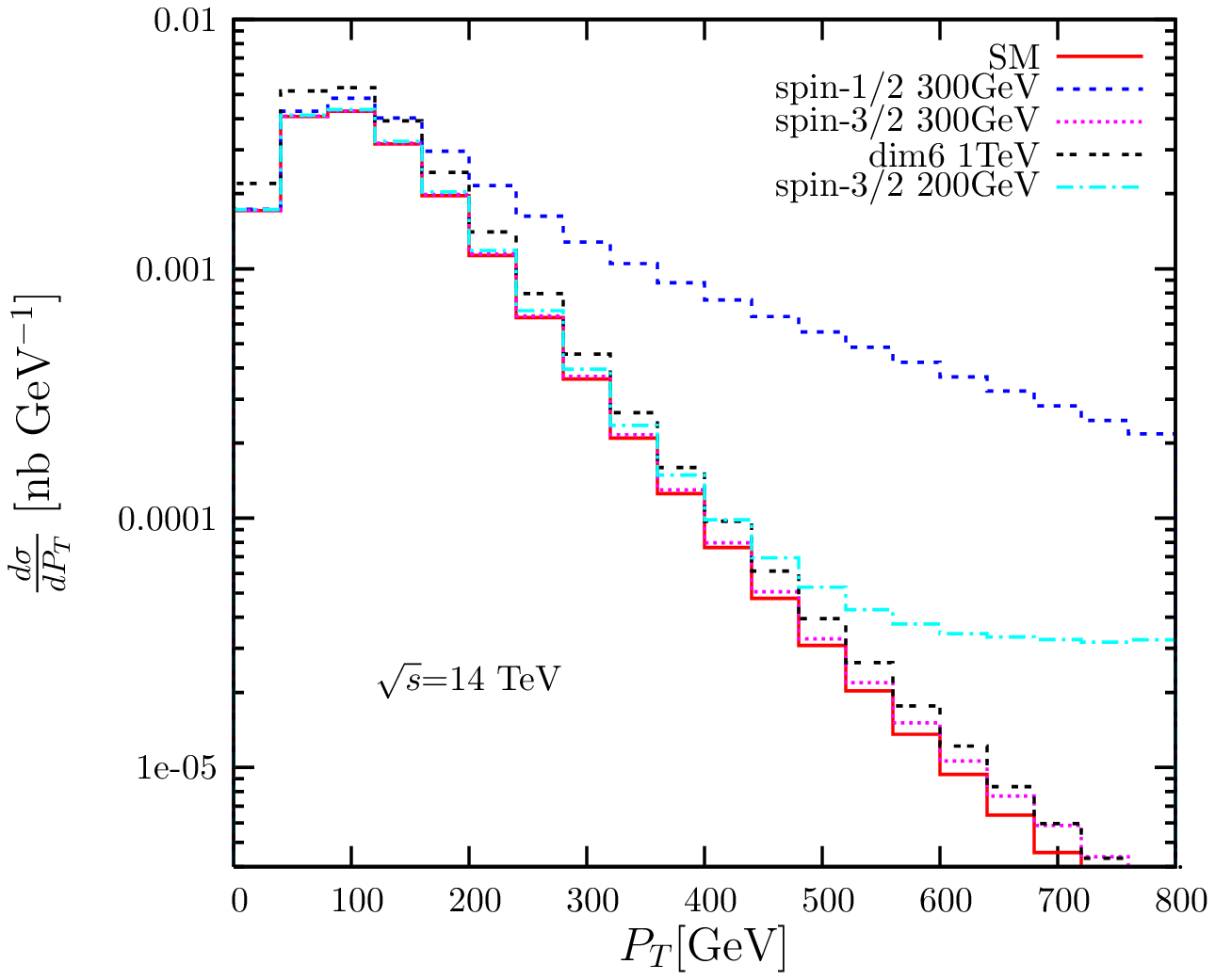}
\end{minipage}
\hspace{0.5cm}
\begin{minipage}[b]{0.5\linewidth}
\centering
\includegraphics[trim=1.5cm 0 0 0 ,scale=0.6]{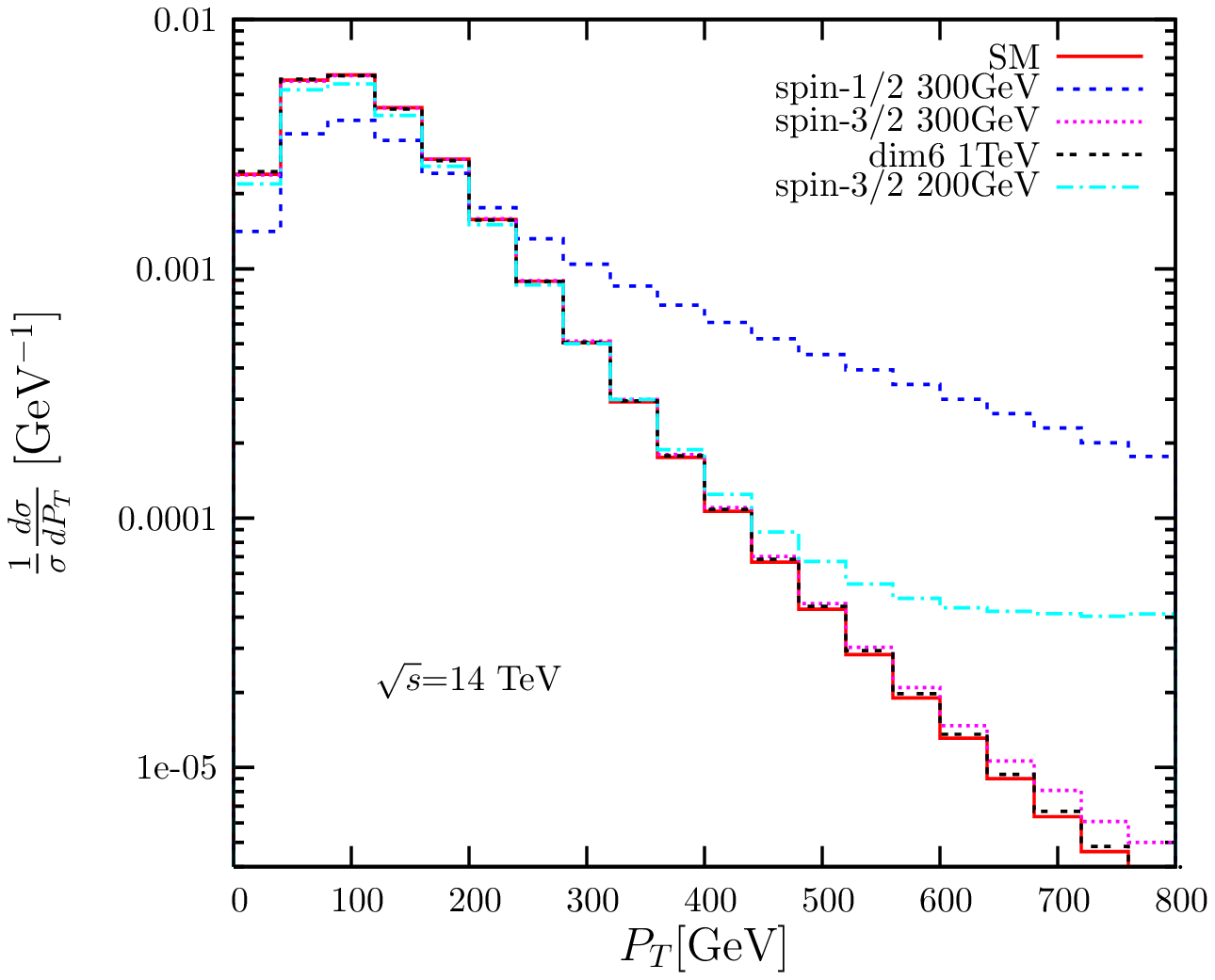}
\end{minipage}
\caption{Unnormalised and normalised differential cross section for transverse top momentum.}
\label{ptnorm2}
\end{figure}

We obtain the differential distributions at the LHC at 14~TeV for the top pair invariant mass and the transverse momentum of the top quark. We compute the differential distributions and the normalised distributions for a spin-1/2 excited quark of mass 300~GeV, two different masses of a spin-3/2 excited top, the dimension-six operator with $c_{hg}/\Lambda^2$=1~TeV$^{-2}$ and the SM. As a reference the results for the total cross section for the different cases are:

\begin{itemize}
\setlength{\itemsep}{0mm}
 \item SM \,\,\,\,\,\,\, 716~nb
\item Dim-6 \,\,\,896~nb
\item Spin-1/2  $M^*_{1/2}=300$~GeV \,\,\,1229~nb
\item Spin-3/2  $M_{3/2}=300$~GeV \,\,\, 722~nb
\item Spin-3/2  $M_{3/2}=200$~GeV \,\,\, 789~nb
\end{itemize}We have chosen $c_{hg}$ to be positive leading to an increase of the cross section.
Even though the increase in the total $t\bar{t}$ cross section varies between the four scenarios from 1\% to more than 60\% it is interesting to see how this increase distributes between different regions of the phase space. The differential distributions for the invariant mass of the top pair and the transverse momentum are shown in Figs.~\ref{mttunno2} and \ref{ptnorm2}. We also show the normalised distributions for both observables. The use of normalised distributions is often employed to alleviate the effect of non-existing NLO calculations. If NLO corrections distribute uniformly in the range of the observable studied then using the normalised distribution for the LO results accounts for the corrections. This is equivalent to the use of a universal $K$-factor. However we emphasize that this is not always the case and NLO corrections often have strong dependence on the region of phase space considered. This should be the object of further study. 

We see from both sets of distributions that we are faced with two fundamentally different cases. The dimension-six operator gives differential distributions with shapes very similar to the SM prediction. The total cross section is increased but the increase is uniform over the whole range of masses and momentum transfers. When we consider the partonic cross section for $gg\rightarrow t\bar{t}$ as a function of the variable $y=\hat{s}/m^2$ for the SM and the interference of the SM with the operator $\mathcal{O}_{hg}$ we see that the threshold functional behaviour proportional to $\sqrt{y-4}$ for $y\sim 4$ emerges in both cases and the shape is similar which explains the uniform deviation from the SM which we see in the top pair invariant mass distributions. On the contrary, in the case of the excited tops, both spin-1/2 and spin-3/2 tops give an excess at high top pair invariant masses and high momentum transfers. Therefore the two scenarios should be easy to distinguish given a deviation from the SM prediction in the experimental results. We note that our SM normalised distributions don't agree with those in \cite{Degrande:2010kt} but are in agreement with those in \cite{Hioki:2009hm}.

To investigate the importance of the neglected $\Lambda^{-4}$ terms for the dimension-six operator in relation to the comparison plots we use the full results for the chromomagnetic operator given in \cite{Haberl:1995ek,Hioki:2009hm} to calculate the cross section at the LHC and the Tevatron. For $c_{hg}=1$ and $\Lambda=1$ TeV we see an increase of 18\% at the LHC and 7\% at the Tevatron of the NP contribution. Even though the effect on the total contribution to the cross section is controlled it is important to appreciate how the inclusion of higher order terms changes the shape of the distributions. We notice that at 14~TeV including the higher order terms increases the cross section at very high invariant mass and transverse momentum, with the cross section becoming generally harder. However the cross section falls to non-detectable values before the effect becomes significant. The additional contribution is also more central. The effect is less pronounced at 7~TeV as the energy limitation does not allow us to reach the regions where the contribution becomes important. Only results with a high imposed cut on the top pair invariant mass would be significantly affected. In general this means that as the mass of the hypothetical excited top increases and the excess shifts to higher top pair invariant masses the difficulty of both detecting an excess and distinguishing the two scenarios increases significantly. 

We also note that the results we get for the excited quarks differ significantly from other resonant effects which would give a peak in the invariant mass distribution at the mass of the resonance, as discussed in  \cite{Frederix:2007gi} for masses in the range accessible given the collider energy\footnote{Although a heavy SM Higgs decaying to $t\bar{t}$ would give a signal of exactly this type, the size of the effect is very small. For example, $\sigma(H\rightarrow t\bar{t})/\sigma(t\bar{t})\sim 10^{-3} (10^{-5})$ for $M_H=500$ GeV (1 TeV).}. However for wide resonance masses which are beyond the reach of the collider we would only see an excess at high energies which would mimic the rise due to the excited tops.

\begin{figure}[h]
\begin{minipage}[b]{0.5\linewidth}
\centering
\includegraphics[trim=1.0cm 0 0 0 ,scale=0.6]{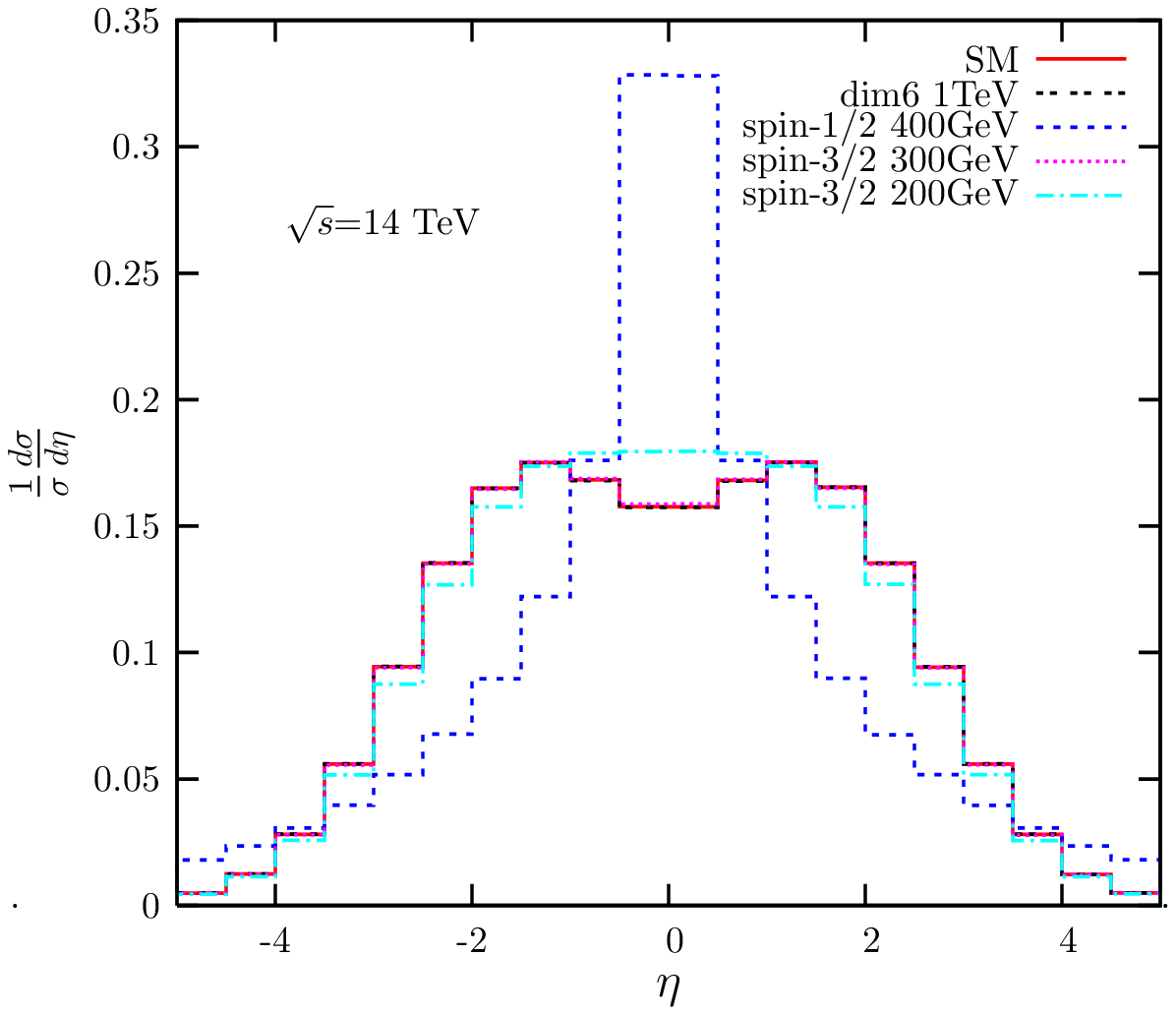}
\caption{Normalised rapidity distribution for the total cross section.}
\label{etanorm}
\end{minipage}
\hspace{0.5cm}
\begin{minipage}[b]{0.5\linewidth}
\centering
\includegraphics[trim=1.6cm 0 0 0 ,scale=0.6]{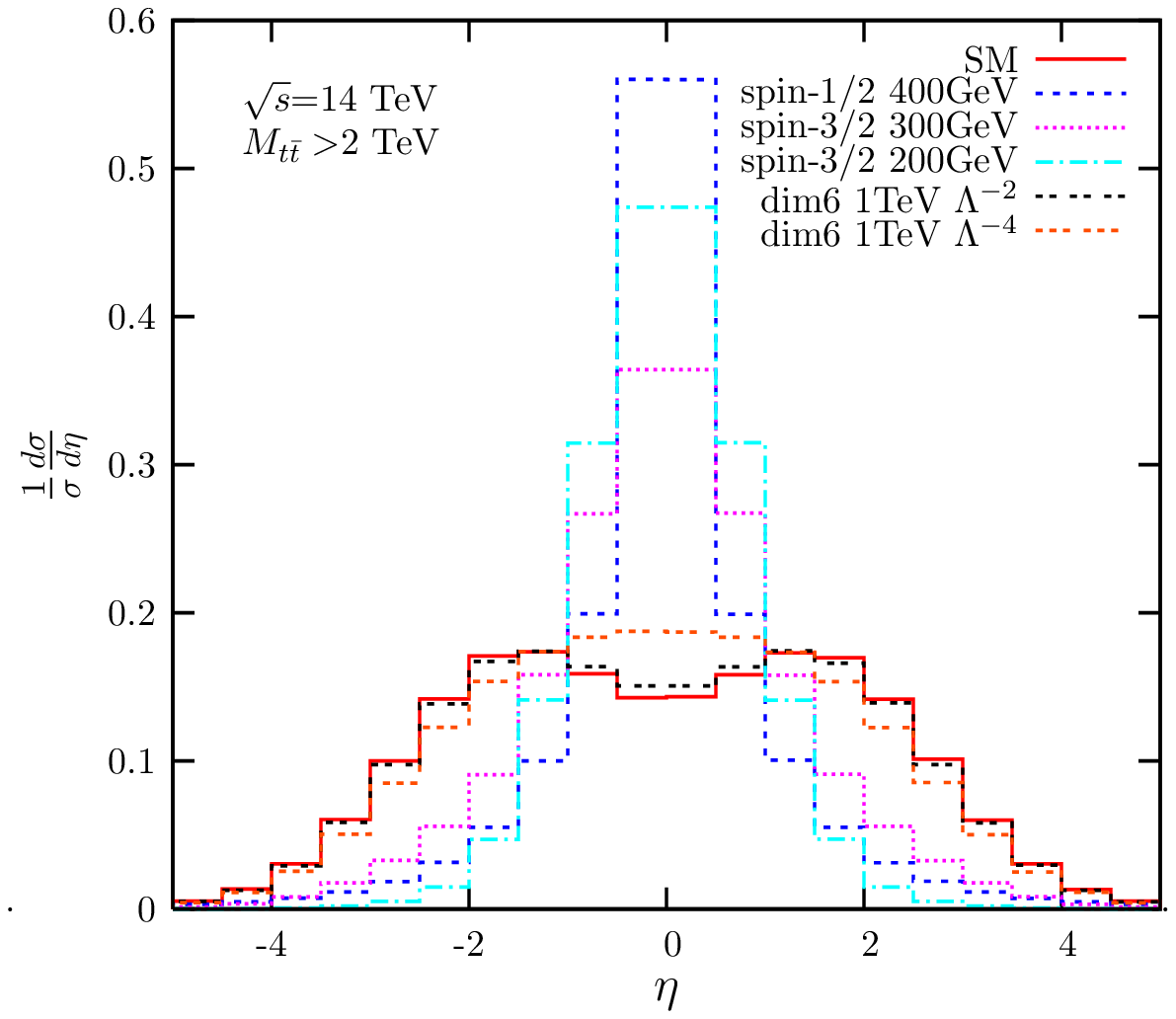}
\caption{Normalised rapidity distribution with an imposed cut of 2~TeV.}
\label{etanormcut}
\end{minipage}
\end{figure}

We also study the rapidity distribution of the top quarks. In all cases we expect a symmetric distribution. The normalised distribution is shown in Fig.~\ref{etanorm}. As the BSM contributions are small the picture is not clear with coincident different scenario lines. To maximise the NP contribution to the total cross section we can impose a cut on the invariant mass of the top quark pair. The result for a cut of 2~TeV is shown in Fig.~\ref{etanormcut}. In the second plot it is shown that for the case of excited quarks the distribution is sharply peaked at the centre while the chromomagnetic operator results in a distribution which remains identical to the SM. As in this case there is an imposed cut on the top pair mass we also show that the inclusion of $\Lambda^{-4}$ terms modifies the prediction for the chromomagnetic operator by enhancing the cross section in the central region.

The corresponding plots for 7 TeV are shown below in Figs.~\ref{mtt7}-\ref{eta7}. The distributions exhibit the same features as for 14~TeV. However the cross sections are generally smaller and the effect of the BSM contributions is less pronounced as we are significantly limited by the CoM energy. The cross section in the region of high invariant mass and transverse momentum where the excess is expected for the excited states is heavily suppressed by the small PDF values for large momentum fractions.  
\begin{figure}[h]
\begin{minipage}[b]{0.5\linewidth}
\centering
\includegraphics[trim=1.5cm 0 0 0 ,scale=0.6]{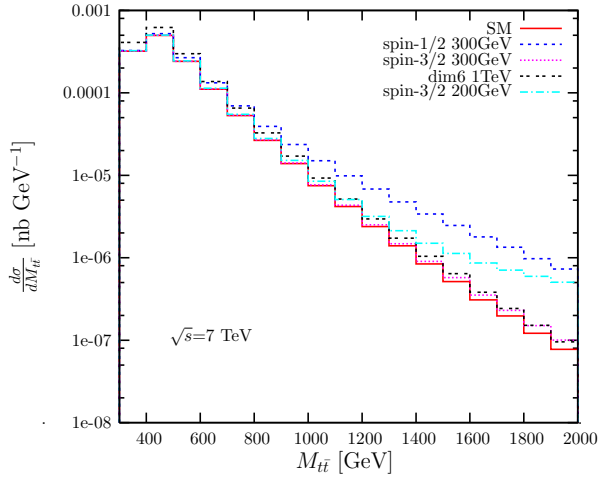}
\end{minipage}
\hspace{0.5cm}
\begin{minipage}[b]{0.5\linewidth}
\centering
\includegraphics[trim=1.6cm 0 0 0 ,scale=0.6]{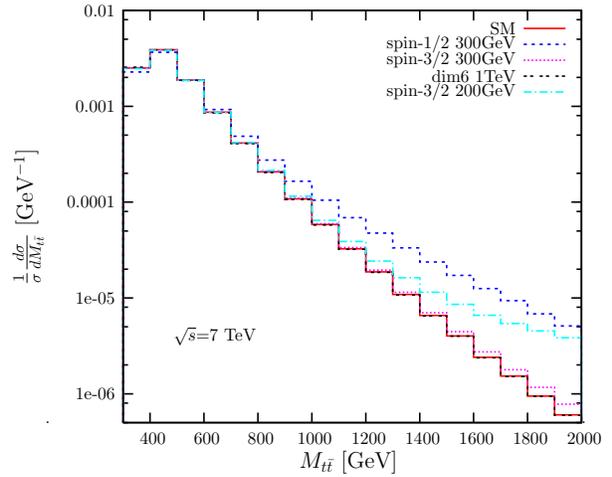}
\end{minipage}
\caption{Unnormalised and normalised differential cross sections for $M_{t\bar{t}}$ at 7 TeV.}
\label{mtt7}
\end{figure}

\begin{figure}[h]
\begin{minipage}[b]{0.5\linewidth}
\centering
\includegraphics[trim=1.9cm 0 0 0 ,scale=0.6]{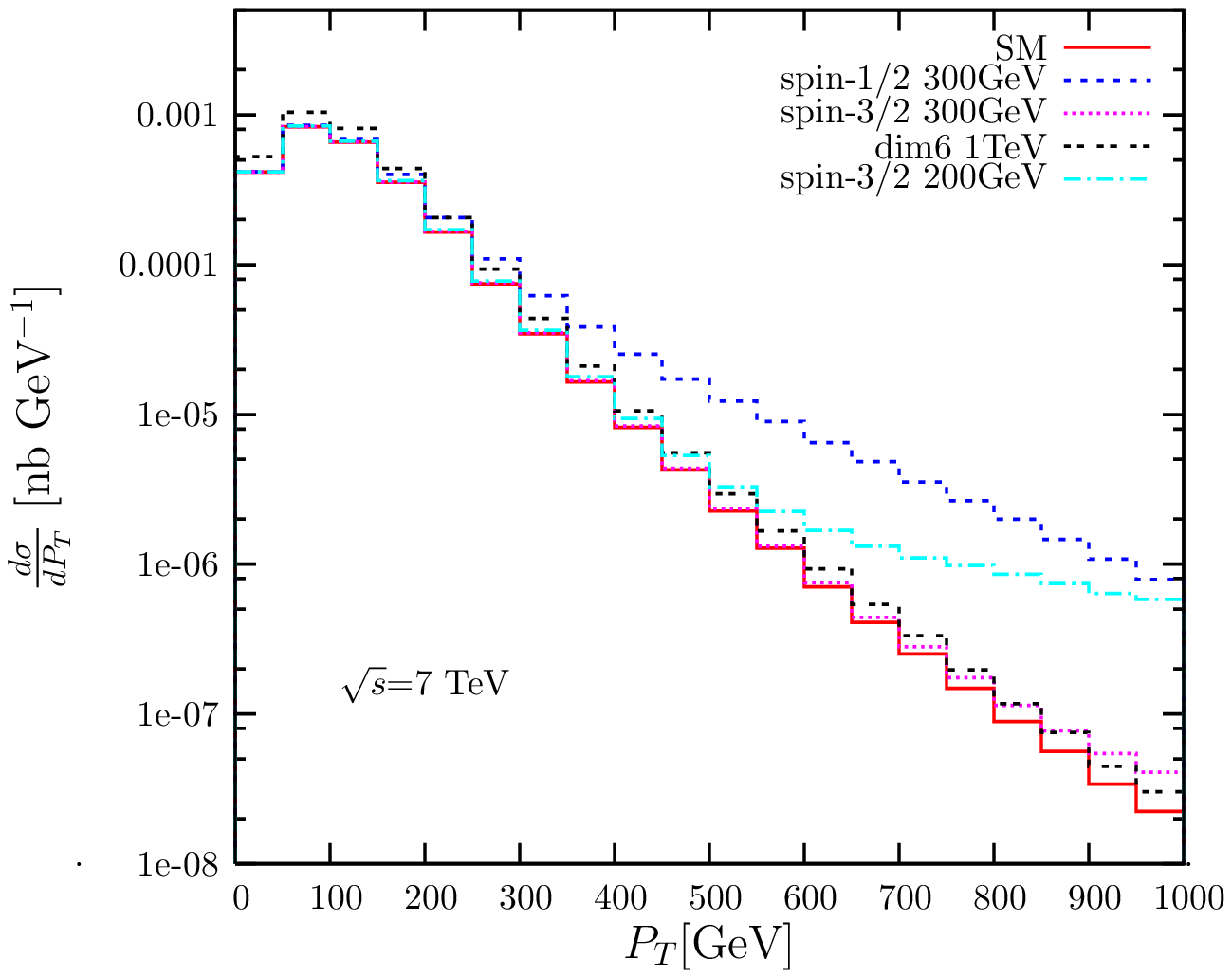}
\end{minipage}
\hspace{0.5cm}
\begin{minipage}[b]{0.5\linewidth}
\centering
\includegraphics[trim=1.7cm 0 0 0 ,scale=0.6]{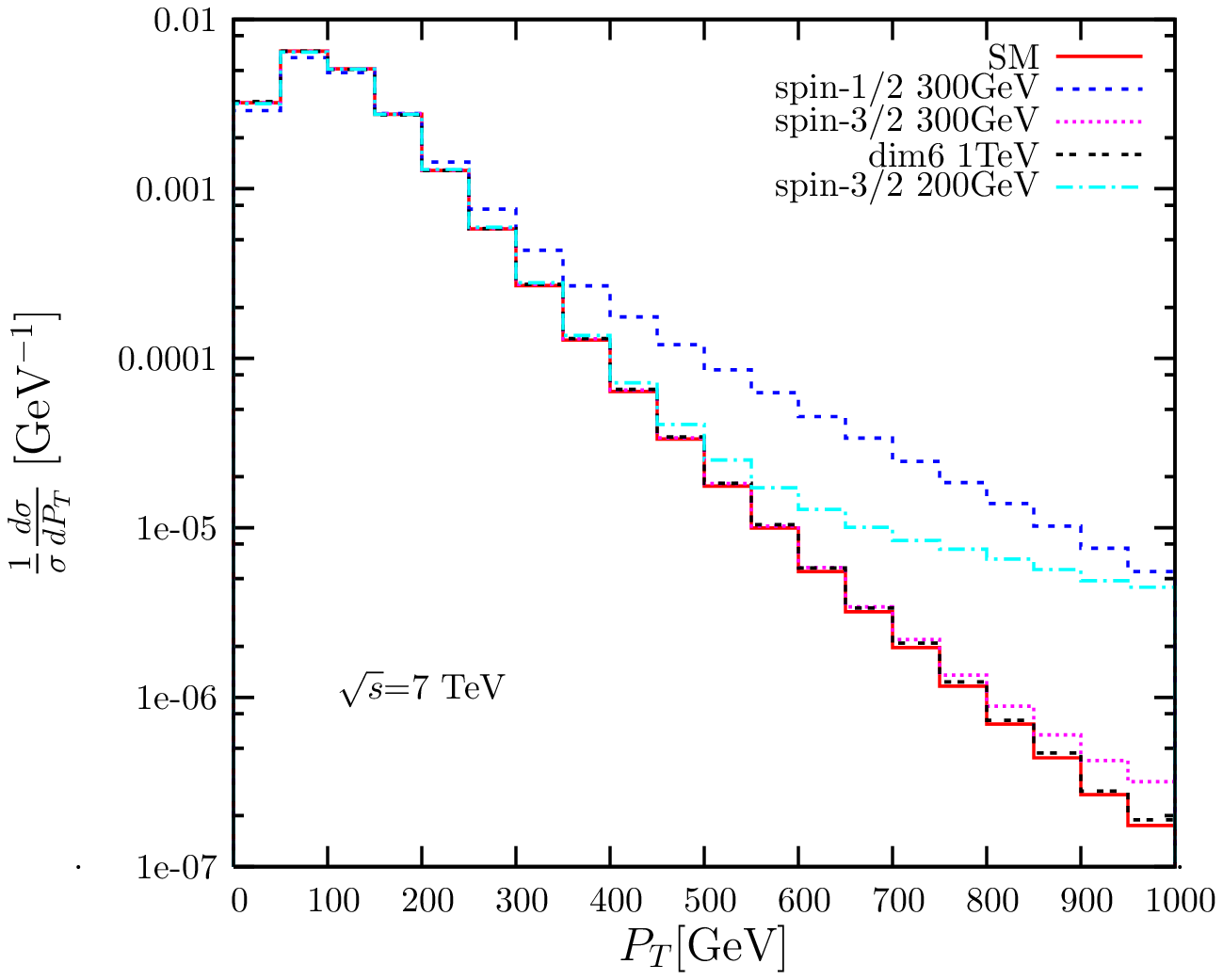}
\end{minipage}
\caption{Unnormalised and normalised differential distributions for transverse top momentum.}
\label{pt7}
\end{figure}
\begin{figure}[h]
\begin{minipage}[b]{0.5\linewidth}
\centering
\includegraphics[trim=1.0cm 0 0 0 ,scale=0.6]{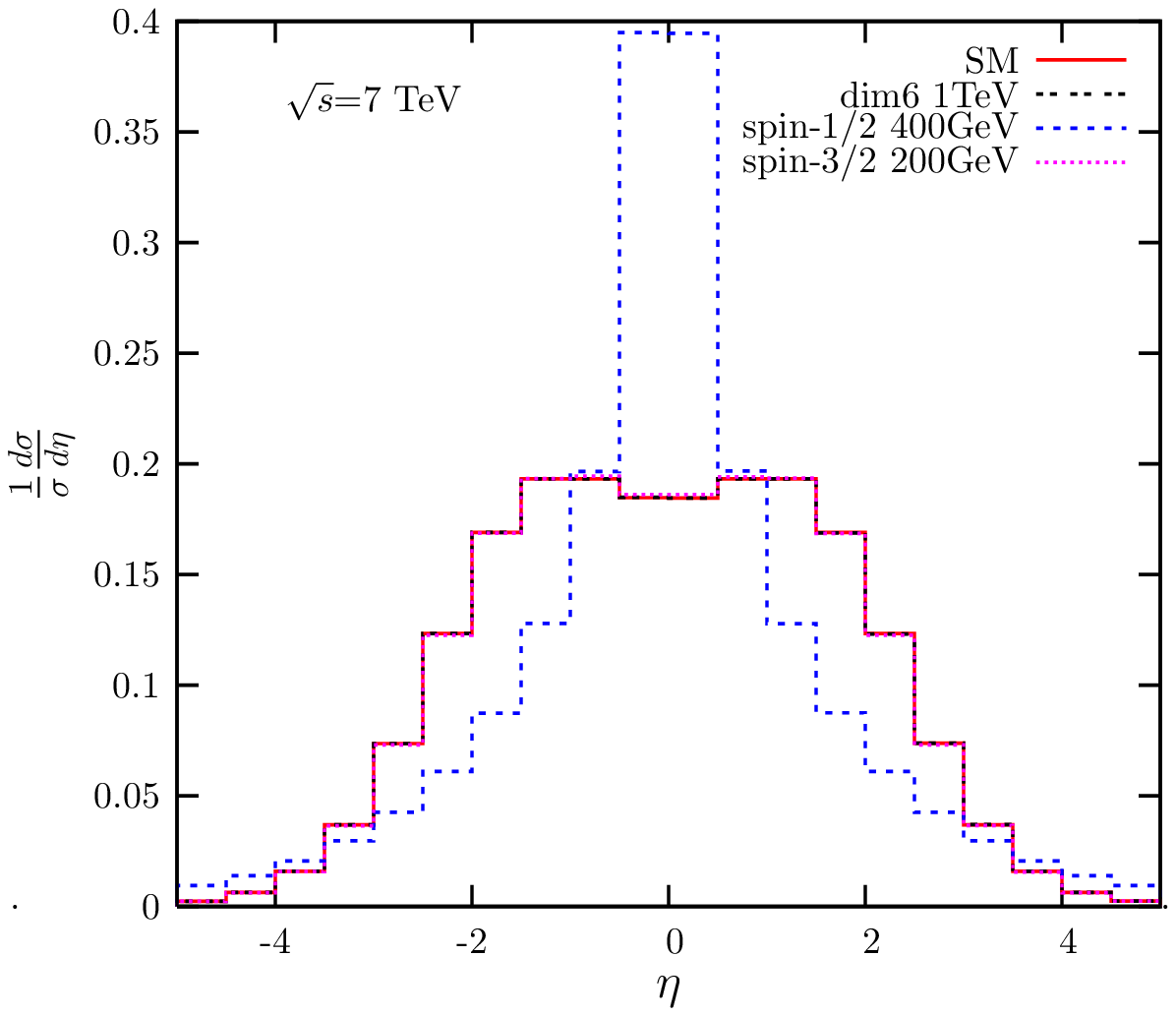}
\end{minipage}
\hspace{0.5cm}
\begin{minipage}[b]{0.5\linewidth}
\centering
\includegraphics[trim=1.1cm 0 0 0 ,scale=0.6]{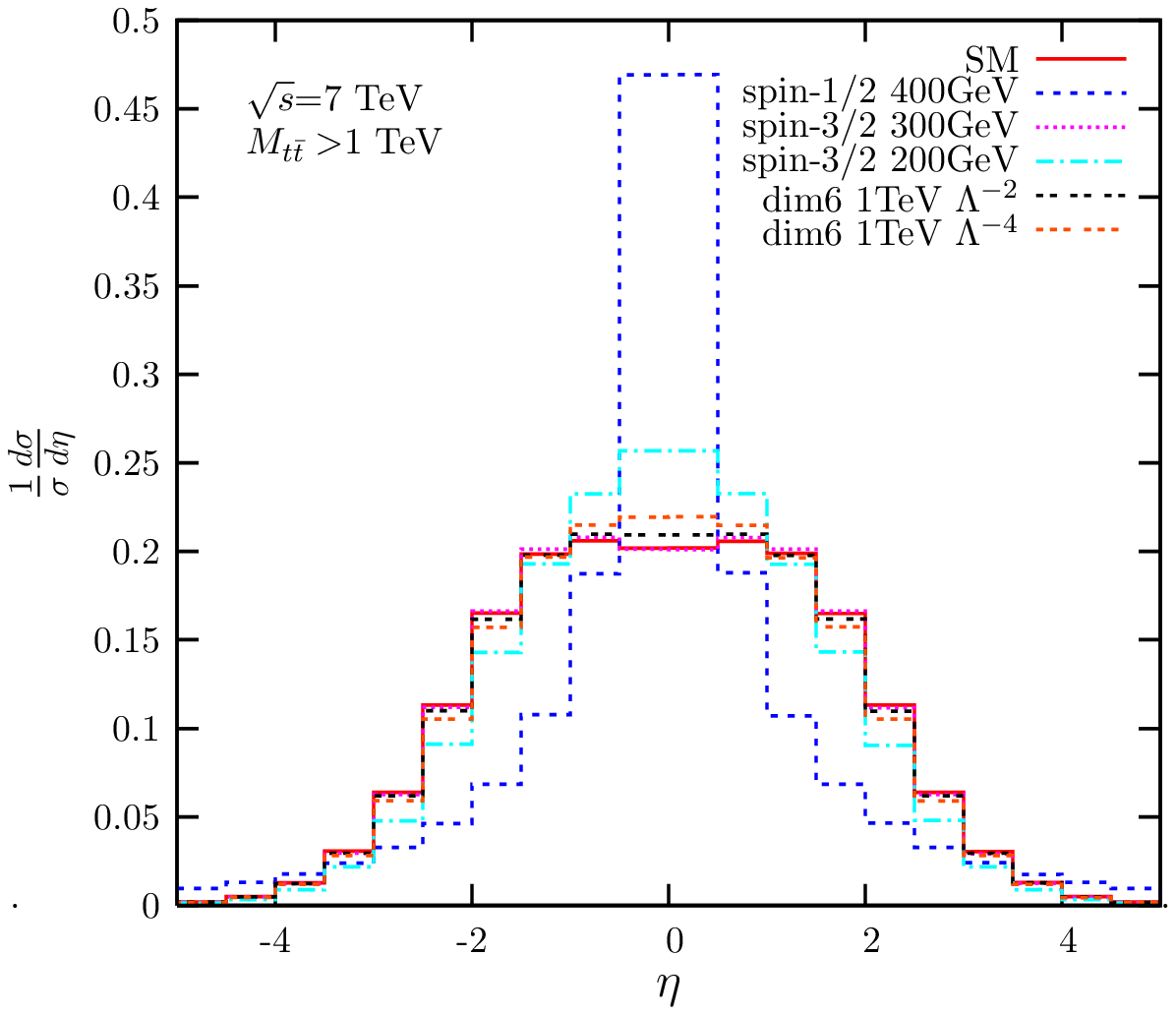}
\end{minipage}
\caption{Normalised rapidity distribution with no cut and an imposed cut of 1~TeV on the invariant mass of the top quark pair.}
\label{eta7}
\end{figure}
\section{Conclusions}
We have studied the effect of spin-3/2 top quark excitations originating in string theory inspired scenarios with warped extra dimensions on the top quark pair production at the LHC. We have investigated the deviation from the SM prediction as a function of the mass of this new state and the off-shell parameter at both 7~TeV relevant for current searches and for 14~TeV as the design LHC energy. By studying the appropriate distributions we have shown that the excited top will give an excess of events in the region of high top pair invariant mass and transverse top momentum. The rapidity distribution of the top quarks is found to be more central than the SM prediction. 

We compared our results with predictions for a hypothetical spin-1/2 excitation and the chromomagnetic operator which gives a uniformly distributed excess above the SM prediction. We have identified the differences between the differential distributions which will help us classify the origin of any future observed excess in top pair events. Of course we must keep in mind that these are tree level calculations and the effect of NLO corrections is expected to affect the results. However as both the final state particles and the colour flow in the NP diagrams are the same as for the SM we might expect the ratio of the BSM results to the SM to be stable with respect to NLO corrections.  

Moreover, extra care is needed in the study of spin-3/2 interactions as they violate unitarity with cross sections rising with energy. A cut-off is introduced and the effect on the reach and exclusion region that can be set by the LHC has been discussed in the light of the measurements undertaken. The current combined results for the top pair cross section at the LHC taking into account different decay channels are: $154\pm {17} \pm 6$(lum) pb from CMS \cite{Chatrchyan:2011yy} and $176\pm 5$(stat)$\pm ^{13}_{10}$(syst)$\pm 7$(lum) from ATLAS\cite{ATLAS}. In terms of the theoretical predictions the main source of uncertainty remains the choice of PDF set, $\alpha_s$ value and renormalisation and factorisation scale as discussed in \cite{Watt:2011kp} where the NLO $t\bar{t}$ cross section at 7~TeV is calculated to vary between 140 and 170 pb for a selection of publicly available PDF sets and their choice of $\alpha_S(M_{Z})$. Given the current accuracy of the top pair cross section measurement at the LHC a narrow region of parameter space is already excluded. For a wider exclusion region one needs to combine these results with the results of searches for single and pair production of the spin-3/2 top which can also be used to set a limit on the mass of this new state. With the increase of the LHC running energy to 14~TeV more definite conclusions can be drawn about the existence of these excitations.

\acknowledgments{E.V. acknowledges financial support from the UK Science and Technology Facilities Council.}


\begin{thebibliography}{99}
\bibitem{Aad:2011yb}
  G.~Aad {\it et al.} [ ATLAS Collaboration ],
    [arXiv:1108.3699 [hep-ex]].

\bibitem{Chatrchyan:2011nb}
  S.~Chatrchyan {\it et al.} [ CMS Collaboration ],
  JHEP {\bf 1107}, 049 (2011).
  [arXiv:1105.5661 [hep-ex]].

\bibitem{Chatrchyan:2011ew}
  S.~Chatrchyan {\it et al.} [ CMS Collaboration ],
   [arXiv:1106.0902 [hep-ex]].

\bibitem{Chatrchyan:2011yy}
  S.~Chatrchyan {\it et al.} [ CMS Collaboration ],
   [arXiv:1108.3773 [hep-ex]].

\bibitem{Abazov:2008ny}
  V.~M.~Abazov {\it et al.} [ D0 Collaboration ],
  Phys.\ Lett.\  {\bf B668}, 98-104 (2008).
 [arXiv:0804.3664 [hep-ex]].

\bibitem{Aaltonen:2009qu}
  T.~Aaltonen {\it et al.} [ CDF Collaboration ],
  Phys.\ Rev.\ Lett.\  {\bf 103}, 041801 (2009).
  [arXiv:0902.3276 [hep-ex]].
\bibitem{Aaltonen:2011ts}
  T.~Aaltonen {\it et al.} [ The CDF Collaboration ],
    [arXiv:1107.5063 [hep-ex]].
\bibitem{ATLASres}
ATLAS Collaboration, ATLAS-CONF-2011-123.
\bibitem{CMSres}
CMS Collaboration, CMS-PAS-TOP-10-007.

\bibitem{Aaltonen:2011kc}
  T.~Aaltonen {\it et al.} [ CDF Collaboration ],
  Phys.\ Rev.\  {\bf D83}, 112003 (2011).
  [arXiv:1101.0034 [hep-ex]].

\bibitem{Abazov:2011rq}
  V.~M.~Abazov {\it et al.} [ D0 Collaboration ],
    [arXiv:1107.4995 [hep-ex]].

\bibitem{Frederix:2007gi}
  R.~Frederix, F.~Maltoni,
  JHEP {\bf 0901}, 047 (2009).
  [arXiv:0712.2355 [hep-ph]].

\bibitem{Barger:2006hm}
  V.~Barger, T.~Han, D.~G.~E.~Walker,
  Phys.\ Rev.\ Lett.\  {\bf 100}, 031801 (2008).
  [hep-ph/0612016].

\bibitem{Kumar:2009vs}
  K.~Kumar, T.~M.~P.~Tait, R.~Vega-Morales,
  JHEP {\bf 0905}, 022 (2009).
  [arXiv:0901.3808 [hep-ph]].

\bibitem{Jung:2009pi}
  D.~-W.~Jung, P.~Ko, J.~S.~Lee, S.~-h.~Nam,
  Phys.\ Lett.\  {\bf B691}, 238-242 (2010).
  [arXiv:0912.1105 [hep-ph]].

\bibitem{Degrande:2010kt}
  C.~Degrande, J.~-M.~Gerard, C.~Grojean, F.~Maltoni, G.~Servant,
  JHEP {\bf 1103}, 125 (2011).
  [arXiv:1010.6304 [hep-ph]].
\bibitem{Burges:1983zg}
  C.~J.~C.~Burges, H.~J.~Schnitzer,
  Nucl.\ Phys.\  {\bf B228}, 464 (1983).

\bibitem{Kuhn:1984rj}
  J.~H.~Kuhn, P.~M.~Zerwas,
  Phys.\ Lett.\  {\bf B147}, 189 (1984).
\bibitem{Kuhn:1985mi}
  J.~H.~Kuhn, H.~D.~Tholl, P.~M.~Zerwas,
  Phys.\ Lett.\  {\bf B158}, 270 (1985).

\bibitem{Moussallam:1989nm}
  B.~Moussallam, V.~Soni,
  Phys.\ Rev.\  {\bf D39}, 1883-1891 (1989).
\bibitem{Almeida:1995yp}
  F.~M.~L.~Almeida, Jr., J.~H.~Lopes, J.~A.~Martins Simoes, A.~J.~Ramalho,
  Phys.\ Rev.\  {\bf D53}, 3555-3558 (1996).
  [hep-ph/9509364].
\bibitem{Dicus:1998yc}
  D.~A.~Dicus, S.~Gibbons, S.~Nandi,
 [hep-ph/9806312].
\bibitem{Walsh:1999pb}
  R.~Walsh, A.~J.~Ramalho,
  Phys.\ Rev.\  {\bf D60}, 077302 (1999).
  [hep-ph/9907364].

\bibitem{Cakir:2007wn}
  O.~Cakir, A.~Ozansoy,
  Phys.\ Rev.\  {\bf D77}, 035002 (2008).
  [arXiv:0709.2134 [hep-ph]].
\bibitem{Hassanain:2009at}
  B.~Hassanain, J.~March-Russell, J.~G.~Rosa,
  JHEP {\bf 0907}, 077 (2009).
  [arXiv:0904.4108 [hep-ph]].
\bibitem{Gherghetta:2000qt}
  T.~Gherghetta, A.~Pomarol,
  Nucl.\ Phys.\  {\bf B586}, 141-162 (2000).
  [hep-ph/0003129].
\bibitem{Rarita:1941mf}
  W.~Rarita, J.~Schwinger,
  Phys.\ Rev.\  {\bf 60}, 61 (1941).

\bibitem{Martin:2009iq}
  A.~D.~Martin, W.~J.~Stirling, R.~S.~Thorne, G.~Watt,
  Eur.\ Phys.\ J.\  {\bf C63}, 189-285 (2009).
  [arXiv:0901.0002 [hep-ph]].

\bibitem{Pumplin:2002vw}
  J.~Pumplin, D.~R.~Stump, J.~Huston, H.~L.~Lai, P.~M.~Nadolsky, W.~K.~Tung,
  JHEP {\bf 0207}, 012 (2002).
  [arXiv:hep-ph/0201195 [hep-ph]].

\bibitem{Georgi:1978kx}
  H.~M.~Georgi, S.~L.~Glashow, M.~E.~Machacek, D.~V.~Nanopoulos,
  Annals Phys.\  {\bf 114}, 273 (1978).

\bibitem{Vermaseren:2000nd}
  J.~A.~M.~Vermaseren,
  arXiv:math-ph/0010025.
\bibitem{Pukhov:2004ca}
  A.~Pukhov,
 [hep-ph/0412191].
\bibitem{Semenov:2010qt}
  A.~Semenov,
  [arXiv:1005.1909 [hep-ph]].
\bibitem{Amsler:2008zzb}
  C.~Amsler {\it et al.} [ Particle Data Group Collaboration ],
  Phys.\ Lett.\  {\bf B667}, 1-1340 (2008).
\bibitem{Nason:1987xz}
  P.~Nason, S.~Dawson, R.~K.~Ellis,
  Nucl.\ Phys.\  {\bf B303}, 607 (1988).

\bibitem{Beenakker:1988bq}
  W.~Beenakker, H.~Kuijf, W.~L.~van Neerven, J.~Smith,
  Phys.\ Rev.\  {\bf D40}, 54-82 (1989).
\bibitem{Langenfeld:2009wd}
  U.~Langenfeld, S.~Moch, P.~Uwer,
  Phys.\ Rev.\  {\bf D80}, 054009 (2009).
  [arXiv:0906.5273 [hep-ph]].
\bibitem{Cacciari:2008zb}
  M.~Cacciari, S.~Frixione, M.~L.~Mangano, P.~Nason, G.~Ridolfi,
  JHEP {\bf 0809}, 127 (2008).
  [arXiv:0804.2800 [hep-ph]].
\bibitem{Kidonakis:2008mu}
  N.~Kidonakis, R.~Vogt,
  Phys.\ Rev.\  {\bf D78}, 074005 (2008).
  [arXiv:0805.3844 [hep-ph]].
\bibitem{Kidonakis:2010dk}
  N.~Kidonakis,
  Phys.\ Rev.\  {\bf D82}, 114030 (2010).
  [arXiv:1009.4935 [hep-ph]].
\bibitem{Kidonakis:2011ca}
  N.~Kidonakis, B.~D.~Pecjak,
   [arXiv:1108.6063 [hep-ph]].

\bibitem{Baur:1989gk}
  U.~Baur, E.~L.~Berger,
  Phys.\ Rev.\  {\bf D41}, 1476 (1990).
  


\bibitem{Benmerrouche:1989uc}
  M.~Benmerrouche, R.~M.~Davidson, N.~C.~Mukhopadhyay,
  Phys.\ Rev.\  {\bf C39}, 2339-2348 (1989).

\bibitem{Pleier:2008ig}
  M.~-A.~Pleier,
  Int.\ J.\ Mod.\ Phys.\  {\bf A24}, 2899-3037 (2009).
  [arXiv:0810.5226 [hep-ex]].

 \bibitem{CMS}
 CMS Collaboration, CMS-PAS-TOP-11-014.
 \bibitem{Baur:1989kv}
   U.~Baur, M.~Spira, P.~M.~Zerwas,
   Phys.\ Rev.\  {\bf D42}, 815-824 (1990).


\bibitem{Buchmuller:1985jz}
  W.~Buchmuller, D.~Wyler,
  Nucl.\ Phys.\  {\bf B268}, 621 (1986).

\bibitem{Atwood:1994vm}
  D.~Atwood, A.~Kagan, T.~G.~Rizzo,
  Phys.\ Rev.\  {\bf D52}, 6264-6270 (1995).
  [hep-ph/9407408].

\bibitem{Cheung:1995nt}
  K.~-m.~Cheung,
  Phys.\ Rev.\  {\bf D53}, 3604-3615 (1996).
  [hep-ph/9511260].
\bibitem{Hioki:2009hm}
  Z.~Hioki and K.~Ohkuma,
  Eur.\ Phys.\ J.\  C {\bf 65}, 127 (2010)
  [arXiv:0910.3049 [hep-ph]].

\bibitem{Haberl:1995ek}
  P.~Haberl, O.~Nachtmann, A.~Wilch,
  Phys.\ Rev.\  {\bf D53}, 4875-4885 (1996).
  [hep-ph/9505409].
\bibitem{ATLAS}
ATLAS Collaboration, ATLAS-CONF-2011-108.
%
\bibitem{Watt:2011kp}
  G.~Watt,
  JHEP {\bf 1109}, 069 (2011).
  [arXiv:1106.5788 [hep-ph]].


\end{thebibliography}
\end{document}